\documentclass[review]{elsarticle}
\usepackage{lineno,hyperref}
\modulolinenumbers[5]
\journal{Journal of \LaTeX\ Templates}

\usepackage{url}
\usepackage{mathrsfs}
\usepackage{caption}
\usepackage{graphicx}
\usepackage{epstopdf}
\usepackage{subfigure}
\usepackage{float}
\usepackage{makecell}
\usepackage{multirow}
\usepackage[table,xcdraw]{xcolor}
\usepackage{bm}

\begin{document}

\begin{frontmatter}

\title{Temporal Gravity Model for Important Nodes Identification in Temporal Networks}





\author[address1]{Jialin Bi}
\ead{jialinbi@mail.sdu.edu.cn}
\author[address1]{Ji Jin}
\ead{mincjinji@gmail.com}
\author[address1,address2]{Cunquan Qu\corref{mycorrespondingauthor}}
\cortext[mycorrespondingauthor]{Corresponding author}
\ead{cqqu@sdu.edu.cn}
\author[address3]{Xiuxiu Zhan}
\ead{x.zhan@tudelft.nl}
\author[address1,address2]{Guanghui Wang}
\ead{ghwang@sdu.edu.cn}

\address[address1]{School of Mathematics, Shandong University, Jinan, 250100, PR China}
\address[address2]{Data Science Institute, Shandong University, Jinan, 250100, PR China}
\address[address3]{Delft University of Technology, Intelligent Systems, Delft, 2600GA, the Netherlands}

\begin{abstract}
Identifying important nodes is one of the central tasks in network science, which is crucial for analyzing the structure of a network and understanding the  dynamical processes on a network. Most real-world systems are time-varying and can be well represented as temporal networks. Motivated by the classic gravity model in physics,  we propose a temporal gravity model to identify influential nodes in temporal networks. Two critical elements in the gravity model are the masses of the objects and the distance between two objects. In the temporal gravity model, we treat nodes as the objects, basic node properties, such as static and temporal properties, as the nodes' masses. We define temporal distances, i.e., fastest arrival distance and temporal shortest distance, as the distance between two nodes in our model.  We utilize our model as well as the baseline centrality methods on important nodes identification. Experimental results on  ten real-world datasets show that the temporal gravity model outperforms the baseline methods in quantifying node structural influence. Moreover, when we use the temporal shortest distance as the distance between two nodes, our model is robust and performs the best in quantifying node spreading influence compared to the baseline methods.
\end{abstract}

\end{frontmatter}


\section{Introduction}
Network science plays an increasingly significant role in numerous domains, including physics, biology, finance, social science, and so on. Many real-world systems can be well represented as complex networks~\cite{Dame2002Statistical,newman2018networks}.

In networks, nodes may play different roles in network connectivity and dynamical processes, such as epidemic spreading, information diffusion, and opinion formation. For instance, if we remove a node from a network and the network will collapse into disconnected components, this node is important in terms of network connectivity. To simplify, we call this kind of influence as the structural influence in the 
remaining part of this work. On the other hand, a node as the seed of an information (epidemic) spreading that can make the information (epidemic) widely circulated, is called an influential node in terms of spreading processes. Analogously, we denote this type of influence as the spreading influence. For both cases, we call the node is an important node in general. The analysis above suggests the importance of analyzing nodes' roles, which induces the study of important node identification in a network~\cite{morone2015influence,tang2009social,zhang2013identifying,zhan2018coupling}. If the connections in a network are fixed, such representation 
is named as static network representation. In terms of static networks, important node identification methods have been well developed~\cite{lu2016vital}. Generally speaking, we can divide these methods into two classes: structural-based centrality methods~\cite{gao2015measures,qiao2018a}, such as degree~\cite{bonacich1972factoring}, closeness~\cite{sabidussi1966the} and betweenness centrality~\cite{freeman1977a}, and iterative-based centrality methods, such as PageRank~\cite{brin1998the}, HITS~\cite{kleinberg1999authoritative} and SALSA~\cite{lempel2001salsa}.
Inspired by the idea of the gravity law, Ma et al.~\cite{ma2016identifying} proposed two gravity models, i.e., gravity centrality and extended gravity centrality, to identify the influential spreaders on static networks by considering both neighborhood and path information. Li et al.~\cite{li2019identifying} proposed a local gravity model by introducing a truncation radius.

However, the methods mentioned above are restricted to static networks. In daily life, many systems are time-varying~\cite{holme2014analyzing, takaguchi2012importance, holme2016temporal, scholtes2014causality}.Studies have shown that the time order has a significant influence not only on the network structure but also the spreading processes on the networks~\cite{takaguchi2012importance,li2017fundamental}. If we consider when the connections happen in a complex system, we can use 
a temporal network to represent the system~\cite{holme2012temporal,renaud2016guide}. The study of identifying important nodes in temporal networks is much more challenging than that in static networks. In temporal networks~\cite{li2017fundamental}, a node may play different roles at different time, which means that the importance of a node also varies with time. For example, an individual that was very active and posted a lot of messages and information last year may become inactive this year. The individual may not be necessary at all for information that starts to spread this year. Thus, to identify important nodes in temporal networks, we need to consider both the structure properties and the time information. Most of the newly proposed metrics for temporal networks are the extension of static ones~\cite{tang2010characterising,grindrod2011communicability}. Some works integrate a temporal network into a static one or cut a temporal network into a series of static snapshots. The extension centrality metrics generally take the following steps. Firstly, a temporal network is divided into several snapshots based on a time resolution. Each of the snapshots is viewed as a static network. We can compute the centrality score of each node on each snapshot. The overall centrality score of a node is the average over all the scores obtained from the snapshots~\cite{kim2012temporal,qu2019temporal}. This type of methods can better identify important nodes compared to the static centrality methods. However, they may lose temporal information, such as the time order of the contact.

In this paper, we propose a temporal gravity model to identify important nodes in temporal networks. Two main elements in universal gravitation are the masses of the objects and the distance between two objects. 
The central assumption of this work is that the centrality of a node depends on its gravitation to nearby nodes. The nearby nodes are determined by the temporal distance, which means the nearby nodes should be close to the target nodes both in structure and time.
In our model, we use node properties, such as static centrality metrics and their extension to temporal networks, as the masses, and the temporal distance between nodes as the distance. The temporal distance between nodes captures both the structure and the time order of the contacts into consideration. We utilize two ways to define the temporal distance between two nodes, i.e., the fastest arrival distance and the temporal shortest distance.
To explore whether temporal gravity model can identify important nodes, we apply it to identify both structural influential nodes and spreading influential nodes in ten empirical temporal networks. We use network efficiency to quantify nodes' real structural influence. We use susceptible-infected-recovered (SIR) spreading model as the spreading process on a temporal network. The node spreading capacity, which is the spreading range of a spreading process originated from the node, is used to measure node's real spreading influence.  The Kendall correlation between a node's real influence and the importance score derived from a centrality metric is computed. The higher Kendall correlation coefficient indicates the better performance of the centrality method in identifying important nodes. The experiments demonstrate that the temporal gravity model outperforms state-of-the-art centrality 
methods significantly in important nodes identification.

This paper is organized as follows. Section 2 first introduces how to represent a temporal network and the definition of temporal distance between nodes. Then we give a brief description of the static and temporal centrality metrics that will be used as the masses of nodes in our temporal gravity model as well as baseline metrics. In Section 3, we illustrate the temporal gravity model. In Section 4, we introduce the empirical temporal network datasets and show the experimental results of the performance evaluation of the temporal gravity model as well as the baseline metrics in Section 5. In Section 6, we conclude our paper.

\section{Preliminaries}
In this section, we first present some basic definitions of temporal networks, including the representation of a temporal network, the definitions of temporal paths, and the concept of distance in temporal networks. Then, we briefly describe benchmark centrality metrics. The centrality metrics and the temporal distance will be used to propose the temporal gravity model on important nodes identification later.

\subsection{Basic Notations and Definitions}
Let $G^{T}=(V,E^{T})$ be a temporal network on time interval $[1,T]$, which consists of a set $V$ of $N=|V|$ nodes and a set of temporal events $E^{T}$. Each event $e\in E^{T}$ is given by a triad $(v_i,v_j,t)$, indicating that node $v_i$ and node $v_j$ have a contact at time $t$. At each time $t\in [1,T]$, the adjacency matrix is $A_t$, in which $A_t(i,j)=1$ represents nodes $v_i$ and $v_j$ are connected at time $t$, otherwise $A_t(i,j)=0$.

We can generate networks with various scales of time if we choose different time resolution of a network data. For example, the email exchange dataset is usually collected by seconds. By setting the time resolution as one hour, we can represent the data by hours, which means a connection forms between two users if they communicated with each other in an hour. We denote the time resolution as $\Delta t$. Then we can use a temporal network $G^{T}$ with $n=T/\Delta t$ snapshots to represent the dataset. The snapshots are given by $G_1,G_2,\cdots G_n$. If $\Delta t$ is small, the temporal network is with a large number of snapshots. If we choose $\Delta t=T$, we obtain the corresponding static network of $G^{T}$, which is denoted by $G=(V, E)$ . A pair of nodes $v_i,v_j$ is connected by a link $(v_i,v_j)\in E$ if they have at least one contact in $G^{T}$. The adjacency matrix of $G$ is donated as $A$, in which $A(i,j)=1$ if nodes $v_i$ and $v_j$ are connected, otherwise $A(i,j)=0$. It is worth noting that each snapshot of $G^{T}$ can be considered as a static network in $\Delta t$.

\begin{figure}[H]
  \centering
  \includegraphics[width=\textwidth]{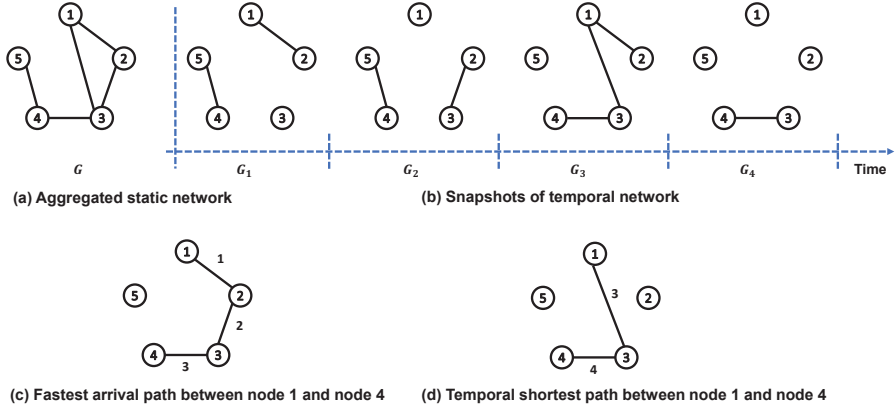}
  \caption{A temporal network $G^{T}=(V,E^{T})$ with 5 nodes and $T=4$ time steps. (a) Aggregated static network $G$. (b) Series of static snapshots $G_1,G_2,G_3, G_4$. (c) Illustration of the fastest arrival path from the start node $1$ to end node $4$. (d) Illustration of the temporal shortest path from the start node $1$ to end node $4$. The corresponding fastest arrival distance and temporal shortest distance are 3 and 2, respectively.}
  \label{Figure:1}
\end{figure}

We show an example of temporal network in Figure~\ref{Figure:1}. Figure \ref{Figure:1}(b) shows a temporal network with 5 nodes and $T=4$ time steps. By setting $\Delta t=1$, the original temporal network contains four snapshots, i.e., $G_1,G_2,G_3, G_4$. In Figure \ref{Figure:1}(a), we give the corresponding aggregated static network $G$.

\subsubsection*{Temporal path}
Given a temporal network $G^{T}=(V,E^{T})$ with $n$ snapshots, a temporal path is a node sequence $P=<v_1,v_{2},\cdots,v_k,v_{k+1}>$, where event $(v_i,v_{i+1},t_i)\in E^{T}$ is the $i$-th temporal event on $P$ for $1\leq i\leq k$ and $t_i\le t_{i+1}$. In this way, $t_1$ is the start time of $P$, denoted as $t_{start}\left(P\right)$, and $ t_k$ is the end time of $P$, denoted as $t_{end}(P)$. We define \emph{the temporal path length} of $P$ as
$l(P)=t_{end}(P)-t_{start}(P)+1$. Given a time internal $[t_a,t_b]$, $v_i$ is the start node and $v_j$ is the end node, $\forall v_i, v_j \in V$. Let ${\textbf{P}}(v_i,v_j,[t_a,t_b])=\{P|P\mbox{ is a temporal path from } v_i\mbox{ to }v_j,\mbox{ s.t. }\\ t_{start}(P)\geq t_a\mbox{ and }t_{end}(P)\leq t_b\}$.
In this paper, we consider two different definitions of temporal paths, i.e., the fastest arrival path and the temporal shortest path~\cite{wu2014path}.

\textit{Fastest arrival path}~\cite{wu2014path}. The fastest arrival path between start node $v_i$ and end node $v_j$ is a temporal path between these two nodes that has the minimum duration counted from $t=1$. That is to say, a fastest arrival path is the path from start node $v_i$ to end node $v_j$  with the minimum elapsed time on a time interval. Therefore, $P\in \textbf{P}(v_i,v_j,[t_a,t_b])$ is a fastest arrival path if $t_{end}(P)= min\{t_{end}(P^\prime)|P^\prime \in 
\textbf{P}(v_i,v_j,[t_a,t_b])\}$. The fastest arrival distance between node $v_i$ and node $v_j$ is the path length of the corresponding fastest arrival path, denoted as $\varphi(v_i,v_j)$.

\textit{Temporal shortest path}~\cite{wu2014path}. The temporal shortest path between the start node $v_i$ and the end node $v_j$ is the path that the overall time needed is the shortest. In other words, $P\in \textbf{P}(v_i,v_j,[t_a,t_b])$ is a temporal shortest path, if $l(P)= min\{l(P^\prime)|P^\prime\in  \textbf{P}(v_i,v_j,[t_a,t_b])\}$. Analogously, the temporal shortest distance between node $v_i$ and node $v_j$ is the path length of the corresponding temporal shortest path, denoted as $\theta(v_i,v_j)$.

We give an example of how to compute these two temporal paths in Figure~\ref{Figure:1} (c) and (d). The fastest path from node $1$ to $4$ is $P_1 = <1,2,3,4>$. The fastest arrival distance $l(P_{1})$ between node $1$ and node $4$ is $\varphi(1,4)=3$. The temporal shortest path from node $1$ to node $4$ is $P_2 = <1,3,4>$, with the temporal shortest distance given by $\theta(1,4)=2$. We also compute the fastest arrival distance and temporal shortest distance between the other nodes, as shown in Table~\ref{Table:2} and Table~\ref{Table:3}, respectively.

\begin{table}[H]
	\caption{The fastest arrival distance between nodes in the temporal network shown in Figure~\ref{Figure:1}. For the nodes that have no fastest arrival paths between them, we denote the distance as $\infty$.}
\begin{tabular}{|p{1cm}<{\centering}|p{1cm}<{\centering}|p{1cm}<{\centering}|p{1cm}<{\centering}|p{1cm}<{\centering}|p{1cm}<{\centering}|}
\hline
\textbf{Node}  & \textbf{1}   & \textbf{2}    & \textbf{3} & \textbf{4} & \textbf{5}    \\ \hline
\textbf{1} & 0    & 1    & 2 & 3 & $\infty$ \\ \hline
\textbf{2} & 1    & 0    & 2 & 3 & $\infty$ \\ \hline
\textbf{3} & 3    & 2    & 0 & 3 & $\infty$ \\ \hline
\textbf{4} & $\infty$ & $\infty$ & 3 & 0 & 1    \\ \hline
\textbf{5} & $\infty$ & $\infty$ & 3 & 1 & 0    \\ \hline
\end{tabular}
\centering
\label{Table:2}
\end{table}

\begin{table}[H]
\caption{ {The temporal shortest distance between nodes in the temporal network given in Figure~\ref{Figure:1}}. For the nodes that have no temporal shortest paths between them, we denote the distance as $\infty$.}
\centering
\begin{tabular}{|p{1cm}<{\centering}|p{1cm}<{\centering}|p{1cm}<{\centering}|p{1cm}<{\centering}|p{1cm}<{\centering}|p{1cm}<{\centering}|}
\hline
\textbf{Node}   & \textbf{1}   & \textbf{2}    & \textbf{3} & \textbf{4} & \textbf{5}    \\ \hline
\textbf{1} & 0    & 1    & 1 & 2 & $\infty$ \\ \hline
\textbf{2} & 1    & 0    & 1 & 2 & $\infty$ \\ \hline
\textbf{3} & 1    & 1    & 0 & 1 & $\infty$ \\ \hline
\textbf{4} & $\infty$ & $\infty$ & 1 & 0 & 1    \\ \hline
\textbf{5} & $\infty$ & $\infty$ & 2 & 1 & 0    \\ \hline
\end{tabular}
\label{Table:3}
\end{table}

\subsection{Benchmark Centrality Metrics}
\label{Benchmark Centrality Metrics}
In this section, we briefly introduce centrality metrics that were proposed before to identify important nodes. 

\textbf{\textit{Degree centrality}}. The degree of a node is defined on the static network $G$, which counts the number of neighbors of the node~\cite{bonacich1972factoring}.
Degree centrality (DC) of a node is the fraction of nodes it is connected to. The greater the degree of a node, the more important it is. The DC is defined as follows:
\begin{equation}
DC\left(i\right)=\frac{k_i}{N-1},
\end{equation}
$k_i$ is the degree of node $v_i$.

\textbf{\textit{Closeness centrality}}. Closeness centrality (CC)~\cite{sabidussi1966the} measures distance between nodes or in practical terms how quickly a node can communicate with all the other nodes in a network. For a disconnected network, this is calculated by the sum of  the reciprocal of the shortest distance from a given node $v_i$ to all the other nodes in the static network $G$:
\begin{equation}
CC\left(i\right)=\frac{1}{N-1}\sum_{i\neq j}\frac{1}{d_{ij}},
\end{equation}
where $d_{ij}$ is the shortest distance between $v_i$ and $v_j$ in $G$.

\textbf{\textit{Betweenness centrality}}. Betweenness centrality (BC)~\cite{freeman1977a} measures the number of shortest paths passing through node $v_i$ in static network $G$, which is given as follows:
\begin{equation}
BC\left(i\right)=\sum_{s\neq i\neq t}\frac{\sigma_{st}^i}{\sigma_{st}},
\end{equation}
where $\sigma_{st}$ is the total number of shortest paths between node $v_s$ and node $v_t$, and $\sigma_{st}^i$ is the number of shortest paths between node $v_s$ and node $v_t$ through node $v_i$.

\textbf{\textit{PageRank}}. PageRank (PR)~\cite{brin1998the} comes from  the Google web search corporation to measure the importance of web pages from the hyperlink network structure. PR assumes that the importance of a web page (node) is determined by its neighbors and the number of pages (nodes) each neighbor linked to, which is defined on the static network $G$. PR is defined as follows:
\begin{equation}
PR(i)^t=\sum_{j=1}^{N}(a_{ij}\frac{PR(j)^{t-1}}{k_{j}^{out}}),
\end{equation}
where ${k_{j}^{out}}$ is the out degree of node $v_j$, and $a_{ij}$ represents the connection between node $v_i$ and $v_j$. $PR(i)^t$ is the PR value of node $v_i$ at time step $t$. After several iterations, the PR value gradually converges and becomes stable.
We use $PR(i)$ to represent the final PR value of node $v_{i}$.

\textbf{\textit{Gravity model}}. The gravity centrality~\cite{ma2016identifying} of a node $v_i$ is defined on static network $G$ and is given by the following equation:
\begin{equation}
g(i)=\sum_{v_j\in\phi_i}\frac{ks_iks_j}{d_{ij}^2},
\end{equation}
where $ks_i$ is the k-shell index of node $v_i$, and $\phi_i$ is the neighbor set of node $v_i$. $d_{ij}$ is the shortest distance between node $v_i$ and node $v_j$ in $G$.

\textbf{\textit{Local gravity model}}. Local gravity centrality~\cite{li2019identifying}  of a node $v_i$ in static network $G$ is defined as follows:
\begin{equation}
g_{R}(i)=\sum_{d_{ij}\leq R,i\neq j}\frac{k_ik_j}{d_{ij}^2},
\end{equation}
where $k_i$ is the degree of node $v_i$, and $d_{ij}$ is the shortest distance between node $v_i$ and node $v_j$. We use $R$ to truncate the contribution of the high-order node on the centrality value of target node $v_i$. In other words, $d_{ij}\leq R$ means nodes that are within distance $R$ to $v_i$ will contribute to the centrality score $g_{R}(i)$. The truncation actually is a trade-off between the local and global structure while considering a node centrality.

\subsection{Centrality Metrics on Temporal Networks}
\label{Centrality Metrics on Temporal Networks}
As illustrated in Section 2.1, each snapshot of a temporal network can be viewed as a static network. Therefore, we can compute the centrality score of each node in each snapshot. The centrality value of a node in a temporal network is the average centrality score over all the snapshots~\cite{kim2012temporal,pan2011path}. We take degree centrality as an example. For node $v_i$, we can compute the degree centrality 
score of $v_i$ on each snapshot to get a $n$-dimensional sequence of the degree centrality scores. Thus, the degree centrality score of node $v_i$ in a temporal network is the average of the $n$-dimensional sequence. For a temporal network with $n$ snapshots $G_1,G_2,\cdots, G_n$, we compute four centrality values, i.e., PR, DC, CC, BC, for every node according to the procedure mentioned above. For the sake of clarity, we use ${PR}^m$, ${DC}^m$, ${CC}^m$, ${BC}^m$ to represent the corresponding average centrality derived from a temporal network with snapshots. We also compute each node's centrality values on the aggregated network. Similarly, we use ${PR}^s$, ${DC}^s$, ${CC}^s$, ${BC}^s$ to denote the centrality computed on the corresponding aggregated static network $G$.

We show examples of computing the centrality score for each node on the temporal network given in Figure~\ref{Figure:1}. The results are shown in Table~\ref{Table:1}, with 'Averaged' meaning the centrality value is given by the average over snapshots of a temporal network and 'Aggregated' meaning the centrality value is derived from the corresponding static network $G$. From Table~\ref{Table:1}, the centrality scores of the same centrality metric derived from a temporal network by averaging over snapshots and from the corresponding static network actually differ a lot.

\begin{table}[H]
\caption{The centrality scores for each node derived from using temporal snapshots ('Averaged') and from using aggregated static network ('Aggregated'). The temporal network and its corresponding static network are show in Figure~\ref{Figure:1}.
}
\centering
\begin{tabular}{|l|l|r|r|r|r|}
\hline
\rowcolor[HTML]{9B9B9B}
\textbf{Node}               & \textbf{Type}       & \textbf{PageRank} & \textbf{Degree} & \textbf{Closeness} & \textbf{Betweenness} \\ \hline
\multirow{2}{*}{1} & Averaged   & 0.163    & 0.188  & 0.203     & 0.083       \\ \cline{2-6}
                   & Aggregated & 0.192    & 0.500  & 0.571     & 0.000       \\ \hline
\multirow{2}{*}{2} & Averaged   & 0.178    & 0.188  & 0.219     & 0.000       \\ \cline{2-6}
                   & Aggregated & 0.192    & 0.500  & 0.571     & 0.000       \\ \hline
\multirow{2}{*}{3} & Averaged   & 0.250    & 0.250  & 0.266     & 0.083       \\ \cline{2-6}
                   & Aggregated & 0.283    & 0.750  & 0.800     & 0.667       \\ \hline
\multirow{2}{*}{4} & Averaged   & 0.265    & 0.250  & 0.281     & 0.000       \\ \cline{2-6}
                   & Aggregated & 0.213    & 0.500  & 0.667     & 0.500       \\ \hline
\multirow{2}{*}{5} & Averaged   & 0.145    & 0.125  & 0.125     & 0.000       \\ \cline{2-6}
                   & Aggregated & 0.120    & 0.500  & 0.444     & 0.000       \\ \hline
\end{tabular}
\label{Table:1}
\end{table}

\section{Temporal Gravity Model}

The classical gravity law contains the masses of the objects as the numerator and the distance between two objects as the denominator. Gravity models, such as the $k$-shell based gravity model and local gravity model, have shown effectiveness in identifying important nodes in static networks. Inspired by the idea of the gravity law and previous gravity models on static networks, we propose a temporal gravity model to identify important nodes in temporal networks. The key point of the temporal gravity method is that node's importance depends on its temporal distance to other nodes as well as the structural properties of the node.

In \textit{temporal gravity model (TG)}, we use node property as its mass, and the distance between two nodes on a temporal network as their distance. Thus, the node importance of $v_i$ is defined as follows:

\begin{equation}
\label{TG}
TG(i)=\sum_{d_{ij}\le R,i\neq j}\frac{M_iM_j}{d_{ij}^2},
\end{equation}
where $M_i$ is the node property of $v_i$, $d_{ij}$ is the temporal distance between nodes $v_i$ and $v_j$, and $R$ is the truncation radius.

For the node property, we use baseline centrality metrics ${PR}^s$, ${DC}^s$, ${CC}^s$, ${BC}^s$, ${PR}^m$, ${DC}^m$, ${CC}^m$, ${BC}^m$ as the mass $M_i$ of a node $v_i$, respectively . Additionally, we propose two types of node degree in temporal networks, i.e., time degree and distance degree, which are also used as the mass of a node in Eq.~(\ref{TG}), respectively. The definitions of time degree and distance degree are given as follows:
\begin{itemize}
\item Time degree (TD)\\
\indent For a temporal network with $n$ snapshots, i.e., $G_1, G_2, \cdots, G_{n}$, the degrees centrality of node $v_i$ on the snapshots are given by $DC(1), DC(2), \cdots, DC(n)$, respectively. We define the time degree (TD) of node $v_i$ as 
$$
TD(i)=e^{DC(1)}+e^{DC(2)}+\cdots +e^{DC(n)}.
$$

\item Distance degree (DD)\\
For node $v_i$, the temporal distance from node $v_i$ to other reachable nodes is $d_{i1},d_{i2},\cdots, d_{ij},\cdots ,d_{im},(d_{ij}\geq 1,1\le\ j\le m\le N)$. The distance degree of $v_i$ is denoted as:
$$
DD(i)=e^{-(d_{i1}-1)}+e^{-(d_{i2}-1)}+\cdots+e^{-(d_{im}-1)}.
$$

The temporal distance that is used in the definition of distance degree takes two ways, i.e., the fastest arrival distance and temporal shortest distance.
\end{itemize}

For the temporal distance used in the denominator of Eq.~\ref{TG}, we consider two cases, i.e., the fastest arrive distance (FAD) and the temporal shortest distance (STD). For FAD, we use the truncation radius equals to the maximal FAD. As the computation complexity of STD is very high, we consider the truncation radius $R=5$. For simplify, we donate the FAD-based and STD-based \textit{TG} method as \textit{TG-fad} and \textit{TG-std}, respectively.

\begin{figure}[H]
  \centering
  \includegraphics[width=0.9\textwidth]{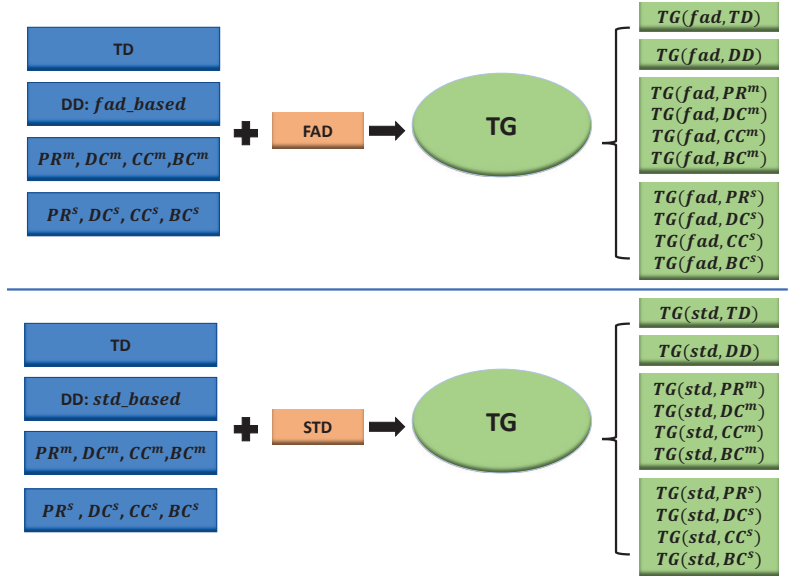}
  \caption{Temporal gravity model with different node properties as node mass and different temporal distances as the distance.}
  \label{Figure:2}%
\end{figure}

For simplification, we write the temporal gravity model in a function-like format $TG(x,y)$, where $x\in\{fad,std\}$ and $y\in\{TD,DD,PR^m,{DC}^m,{CC}^m,{BC}^m\\,PR^s,{DC}^s,{CC}^s,{BC}^s\}$. For example, if we take $DD$ as the mass and $FAD$ as the distance in Eq.~\ref{TG}, then the model is denoted as \textit{TG(fad,DD)}. On the whole, we show a clear diagram of the temporal gravity model via using different forms of mass and distance in Figure~\ref{Figure:2}.

\section{Empirical Networks}
We evaluate the performance of the temporal gravity model on the following temporal empirical network datasets.  Some of the detailed properties of the networks are shown in Table~\ref{Table:4}.

\begin{itemize}
\item High school 2011 (2012,2013) dynamic contact networks~\cite{fournet2014contact,mastrandrea2015contact} (HS2011, HS2012, HS2013). These datasets correspond to the contacts and friendship relations between students in a high school in Marseilles, France.
\item Workplace (WP)~\cite{genois2014data}. This data set contains contacts between employees in an office building in France, from June 24 to July 3, 2013.
\item Haggle~\cite{chaintreau2007impact}. This network contains contacts between people measured by carried wireless devices.
\item Hospital contract (HC)~\cite{vanhems2013estimating}. This dataset contains contacts between patients, contacts between patients and health-care workers (HCWs), and contacts between HCWs in a hospital ward in Lyon, France, from December 6 to December 10, 2010. 
\item Primary school (PS)~\cite{gemmetto2014mitigation}. This data set contains contacts between the children and teachers used in the study published in BMC Infectious Diseases 2014.
\item Hypertext2009 (HT2009)~\cite{isella2011s}. This network contains contacts between the attendees of the ACM Hypertext 2009 conference.
\item Infectious~\cite{isella2011s}. This network contains contacts between people during the exhibition INFECTIOUS: STAY AWAY in 2009 at the Science Gallery in Dublin.
\item SFHH conference (SFHH)~\cite{genois2018can}. This dataset contains contacts between participants in the 2009 SFHH conference in Nice, France.

\end{itemize}
\begin{table}[H]
\caption{Property description of the empirical networks. The number of nodes $N$, the original length of the observation time $T$, time resolution $\Delta t$ (Its unit is hour, recorded as $H$), the number of snapshots $n$, the total number of contacts $\left|C\right|$, and the number of links $\left|E\right|$ in aggregated static network $G$ are shown.}
\centering
\begin{tabular}{|l|l|r|r|r|r|r|r|}
\hline
\rowcolor[HTML]{9B9B9B}
    \textbf{Network}            & \textbf{N}   & \textbf{T}      & \bm{$\Delta t$}     & \textbf{n}   & \bm{$\left|C\right|$}     & \bm{$\left|E\right|$}   \\ \hline
  HS2011  & 126 & 5,609  & H    & 76  & 28,561  & 1,710 \\ \hline
  HS2012  & 180 & 11,273 & H    & 203 & 45,047  & 2,239 \\ \hline
  HS2013  & 327 & 7,375  & H    & 101 & 188,508 & 5,818 \\ \hline
 WP         & 92  & 7,104  & H    & 275 & 9,827   & 755   \\ \hline
 Haggle             & 274 & 15,662 & H    & 96  & 28,244  & 2,899 \\ \hline
 HC & 75  & 9,453  & H    & 97  & 32,424  & 1,139 \\ \hline
 PS     & 242 & 3,100  & 0.5H & 65  & 125,773 & 8,317 \\ \hline
HT2009      & 113 & 5,246  & 0.5H & 118 & 20,818  & 2,196 \\ \hline
Infectious         & 410 & 1,392  & 0.1H & 79  & 17,298  & 2,765 \\ \hline
SFHH               & 403 & 3,509  & 0.5H & 64  & 70,261   & 9,889 \\ \hline
\end{tabular}
\label{Table:4}%
\end{table}

\section{Results}
The centrality metrics illustrated in Section~\ref{Benchmark Centrality Metrics} and~\ref{Centrality Metrics on Temporal Networks} are used as the baseline metrics in contrast to the temporal gravity model. To evaluate the performance of the temporal gravity model as well as the baseline metrics, we propose to use network efficiency and susceptible-infected-recovered (SIR) spreading model on temporal networks as the performance evaluation methods. The network efficiency aims to evaluate a node's role in exchanging information, whereas the SIR spreading model aims to evaluate a node's role in spreading capacity. The node importance score derived from different centrality metrics and the performance evaluation methods, i.e., network efficiency and the SIR spreading model, are compared by using Kendall correlation coefficient $\tau$\footnote{Kendall's Tau~\cite{kendall1938new} is an index measuring the correlation strength between two sequences. The larger Kendall's Tau, the greater the similarity between two sequence. Considering two sequences with $N$ elements, $X$ =$(x_1, x_2,\cdots, x_N)$ and $Y=(y_1, y_2, \cdots, y_N)$. Any pair of two-tuples $(x_i, y_i)$ and $(x_j, y_j)(i\neq j)$ is concordant if both $x_i > x_j$ and $y_i > y_j$ or both $x_i < x_j $ and $y_i < y_j$. It is discordant if $x_i > x_j$ and $y_i<y_j$ or $x_i < x_j$ and $y_i > y_j$. If $x_i = x_j$ or $y_i = y_j$, the pair is neither concordant nor discordant. Kendall's Tau of two sequences $X$ and $Y$ can be calculated as:
$$
\tau=\frac{1}{N(N-1)}\sum_{i\neq j}{sgn(x_i-x_j)sgn(y_i-y_j)}.
$$}. The high value of Kendall correlation coefficient $\tau$ indicates the centrality metric can better identify important nodes in a temporal network and vice verse. In this section, we first define network efficiency on temporal networks and present results of identifying structural influence nodes in temporal networks by using centrality metrics including temporal gravity model. Later on, we compare the temporal gravity model and the baseline centrality metrics on the performance of identifying influential nodes in an SIR spreading process on temporal networks.

\subsection{Performance Evaluation Based on Network Efficiency}
Network efficiency~\cite{latora2001efficient} is defined based on the assumption that the information in a network passes only through the temporal shortest paths. The efficiency of a network measures how efficiently information is exchanged over the network.
We define the network efficiency of a temporal network $G^{T}$ as follows:
\begin{equation}
\varepsilon\left(G^{T}\right)=\frac{1}{N(N-1)}\sum_{v_i\neq v_j\in G^{T}}\frac{1}{d_{ij}},
\end{equation}
where $d_{ij}$ is the temporal distance between node $v_i$ and $v_j$. The temporal distance between two nodes takes two scenarios, i.e., fastest arrival distance and temporal shortest distance.

If we remove a node from a temporal network, it may decrease the network efficiency. Because it may make the network disconnected. Therefore, the reduction of efficiency after nodes' removal is used to measure the importance of nodes in temporal networks~\cite{qu2019temporal}. Larger reduction of efficiency means the node is more important in terms of structure influence.

We use $G^{T}\setminus v_i$ to denote the temporal network after removing node $v_i$ and all the contacts associated with $v_i$. The difference between the network efficiency of $G^{T}$ and the network efficiency of $G^{T}\setminus v_i$ is defined as the importance score of node $v_i$ regarding to network efficiency. The formula is given as follows: 
\begin{equation}
 NE(v_i)=\varepsilon(G^{T})-\varepsilon(G^{T}\setminus v_i).
\end{equation}

Consequently, node efficiency based on fastest arrival distance and temporal shortest distance are denoted as \textit{$NE_{fad}$} and \textit{$NE_{std}$}, respectively.

We show the performance of the temporal gravity model and the baseline centrality metrics in identifying important nodes with regard to network efficiency in Table~\ref{Table:6}, in which we use the fastest arrival distance as the distance in the temporal gravity model in this scenario. The values in Table~\ref{Table:6} show the Kendall correlation coefficient $\tau$ between the node centrality scores derived by corresponding centrality metrics and the node efficiency based on fastest arrival distance. We give the results of ten temporal empirical networks. In general, the temporal gravity model can better identify important nodes than the baseline centrality metrics, i.e., ${PR}^m, {DC}^m, {CC}^m, {BC}^m, {PR}^s, {DC}^s, {CC}^s, {BC}^s$. Specifically, the Kendall correlation coefficient $\tau$ of the temporal gravity model increases by 105.31$\%$ on average compared to the best value of $\tau$ derived from baseline centrality metrics 
across the ten empirical networks. In particular, 
the Kendall correlation coefficient $\tau$ of network \textit{Infectious} increases from 0.19491 to 0.83589 from the best of the baseline metrics to the best of the temporal gravity model, with the improvement of 328.85$\%$.
\begin{table}[H]
\caption{The Kendall correlation coefficient $\tau$ between the node centrality score derived from $NE_{fad}$ and the centrality metrics for ten empirical networks. The best $\tau$ of each network data is emphasized in bold and asterisk. The best $\tau$ of each network data derived from the baseline metrics is emphasized in bold. }
\centering
\resizebox{\textwidth}{!}{
\begin{tabular}{|l|r|r|r|r|r|r|r|r|r|r|}
\hline
\rowcolor[HTML]{9B9B9B}
    & \multicolumn{1}{l|}{\cellcolor[HTML]{9B9B9B}\textbf{HS2011}}    & \multicolumn{1}{l|}{\cellcolor[HTML]{9B9B9B}\textbf{HS2012}}    & \multicolumn{1}{l|}{\cellcolor[HTML]{9B9B9B}\textbf{HS2013}}    & \multicolumn{1}{l|}{\cellcolor[HTML]{9B9B9B}\textbf{WP}}        & \multicolumn{1}{l|}{\cellcolor[HTML]{9B9B9B}\textbf{Haggle}}    & \multicolumn{1}{l|}{\cellcolor[HTML]{9B9B9B}\textbf{HC}}        & \multicolumn{1}{l|}{\cellcolor[HTML]{9B9B9B}\textbf{PS}}        & \multicolumn{1}{l|}{\cellcolor[HTML]{9B9B9B}\textbf{HT2009}}    & \multicolumn{1}{l|}{\cellcolor[HTML]{9B9B9B}\textbf{Infections}} & \multicolumn{1}{l|}{\cellcolor[HTML]{9B9B9B}\textbf{SFHH}}      \\ \hline
$TG(fad,DD)$ & 0.71166                        & 0.68206                        & 0.59523                        & 0.65551                        & 0.79408                        & 0.90788                        & 0.53910                        & 0.77152                        & 0.80034                        & 0.77742                        \\ \hline
$TG(fad,TD)$ & 0.72902                        & 0.70478                        & 0.60500                        & 0.66890                        & 0.79224                        & \textbf{0.91207$^\ast$} & 0.56497                        & 0.77118                        & 0.81588                        & 0.78292                        \\ \hline
$TG(fad,{PR}^m)$  & 0.73206                        & 0.72502                        & 0.63329                        & 0.68323                        & \textbf{0.80519$^\ast$}& 0.88180                        & \textbf{0.62505$^\ast$} & 0.79741                        & \textbf{0.83589$^\ast$} & \textbf{0.82366$^\ast$} \\ \hline
$TG(fad,{PR}^s)$  & 0.73206                        & 0.75233                        & \textbf{0.67115$^\ast$} & 0.74104                        & 0.76653                        & 0.87459                        & 0.49604                        & 0.78729                        & 0.79583                        & 0.77739                        \\ \hline
$TG(fad,{DC}^m)$  & 0.71632                        & 0.69758                        & 0.62841                        & 0.72193                        & 0.72517                        & 0.78883                        & 0.57745                        & 0.72408                        & 0.69442                        & 0.75630                        \\ \hline
$TG(fad,{DC}^s)$  & 0.71606                        & 0.73979                        & 0.66849                        & \textbf{0.75108$^\ast$} & 0.72988                        & 0.85225                        & 0.46723                        & 0.77655                        & 0.71594                        & 0.74235                        \\ \hline
$TG(fad,{CC}^m)$  & 0.73333                        & 0.69944                        & 0.63108                        & 0.73053                        & 0.66157                        & 0.81189                        & 0.52711                        & 0.75411                        & 0.73038                        & 0.79462                        \\ \hline
$TG(fad,{CC}^s)$   & \textbf{0.74375$^\ast$} & \textbf{0.72539$^\ast$} & 0.64233                        & 0.68849                        & 0.80295                        & 0.91063                        & 0.59473                        & \textbf{0.80468$^\ast$} & 0.83317                        & 0.80798                        \\ \hline
$TG(fad,{BC}^m)$    & 0.69973 & 0.65347 & 0.61902 & 0.66444 & 0.41410 & 0.59495 & 0.46867 & 0.62263 & 0.54512 & 0.62040 \\ \hline
$TG(fad,{BC}^s)$  & 0.63378 & 0.62702 & 0.61839 & 0.72241 & 0.34422 & 0.75063 & 0.37945 & 0.66498 & 0.50254 & 0.62421  \\ \hline
${PR}^m$ & 0.41029                        & 0.38423                        & 0.37716                        & 0.30244                        & 0.34642                        & 0.30523                        & 0.33336                        & 0.33028                        & 0.07638                        & 0.42005                        \\ \hline
${DC}^m$ & 0.51531                        & 0.40710                        & 0.41160                        & 0.31372                        & 0.46549                        & 0.32966                        & \textbf{0.36322} & 0.36002                        & 0.09907                        & 0.41263                        \\ \hline
${CC}^m$ & \textbf{0.53041} & \textbf{0.41043} & 0.41014                        & 0.32011                        & 0.39340                        & 0.30306                        & 0.26374                        & \textbf{0.41056 }& 0.16953                        & \textbf{0.45180} \\ \hline
${BC}^m$ & 0.51660                        & 0.34806                        & 0.37426                        & 0.27800                        & 0.37925                        & 0.27712                        & 0.28027                        & 0.27497                        & \textbf{0.19491} & 0.33410                        \\ \hline
${PR}^s$ & 0.46210                        & 0.34364                        & 0.40947                        & 0.39943                        & 0.38846                        & 0.37946                        & 0.22883                        & 0.36315                        & 0.13729                        & 0.36444                        \\ \hline
${DC}^s$ & 0.47604                        & 0.35421                        & \textbf{0.41230 }& \textbf{0.41010 }& \textbf{0.47757} & 0.38060                        & 0.22845                        & 0.36582                        & 0.11060                        & 0.36446                        \\ \hline
${CC}^s$ & 0.44594                        & 0.30643                        & 0.39309                        & 0.40781                        & 0.39506                        & \textbf{0.38671} & 0.22781                        & 0.36740                        & 0.17343                        & 0.36301                        \\ \hline
${BC}^s$& 0.43975                        & 0.27940                        & 0.38872                        & 0.37458                        & 0.33128                        & 0.37225                        & 0.23398                        & 0.35240                        & 0.17433                        & 0.35885                        \\ \hline
\end{tabular}}
\label{Table:6}
\end{table}

\begin{figure}[H]
  \centering
  
  \subfigure{\includegraphics[width=\textwidth]{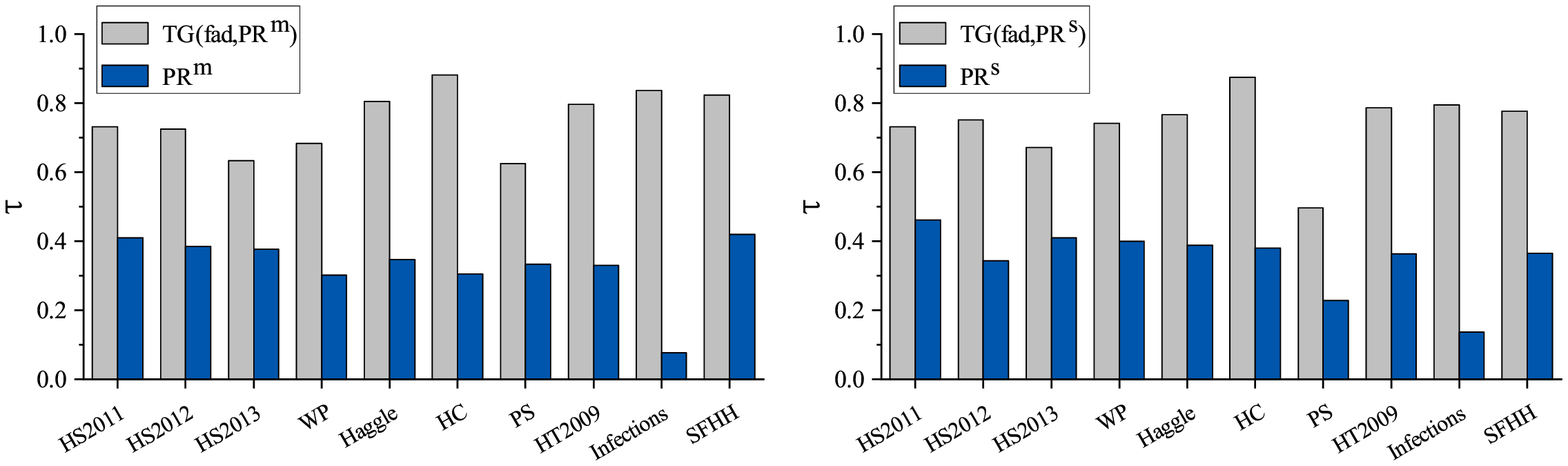}} 

  \subfigure{\includegraphics[width=\textwidth]{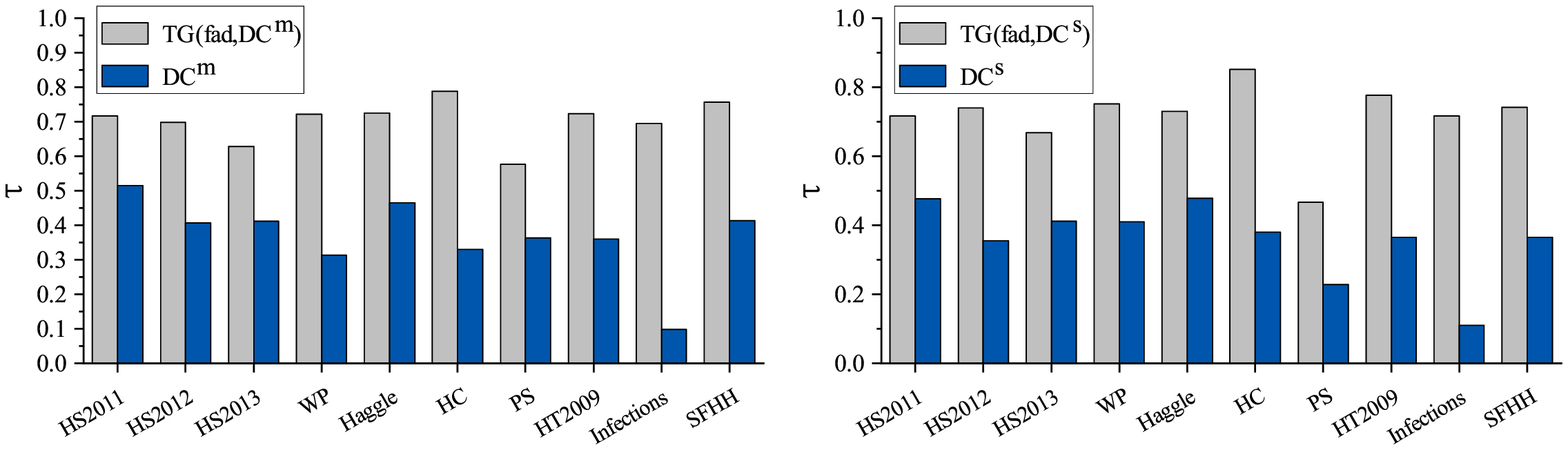}} 

  \subfigure{\includegraphics[width=\textwidth]{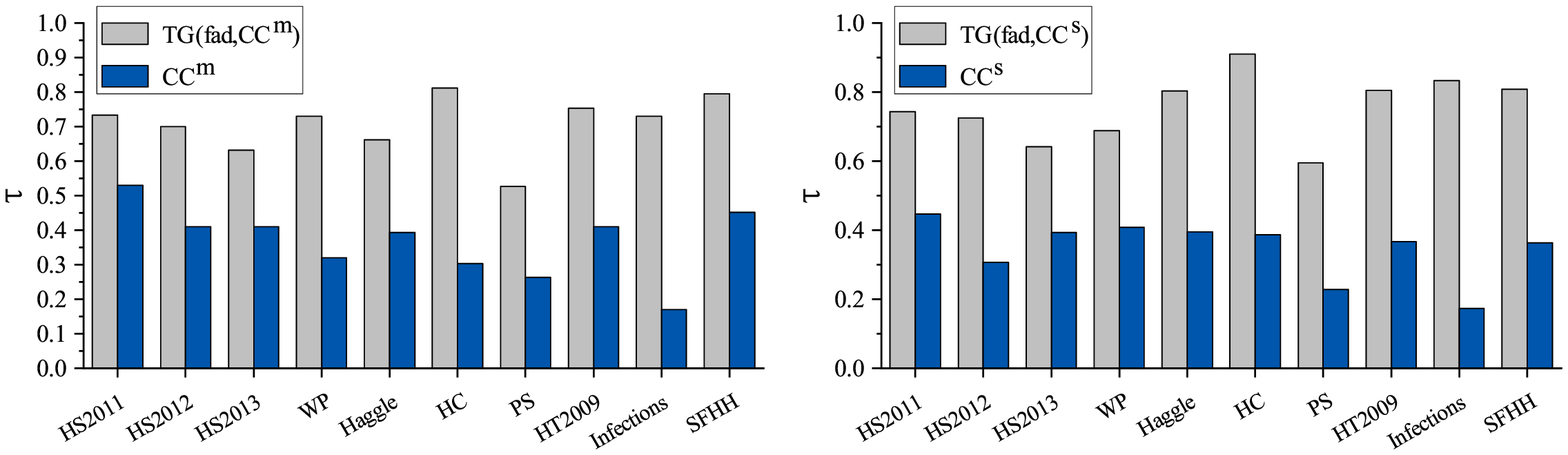}} 

  \subfigure{\includegraphics[width=\textwidth]{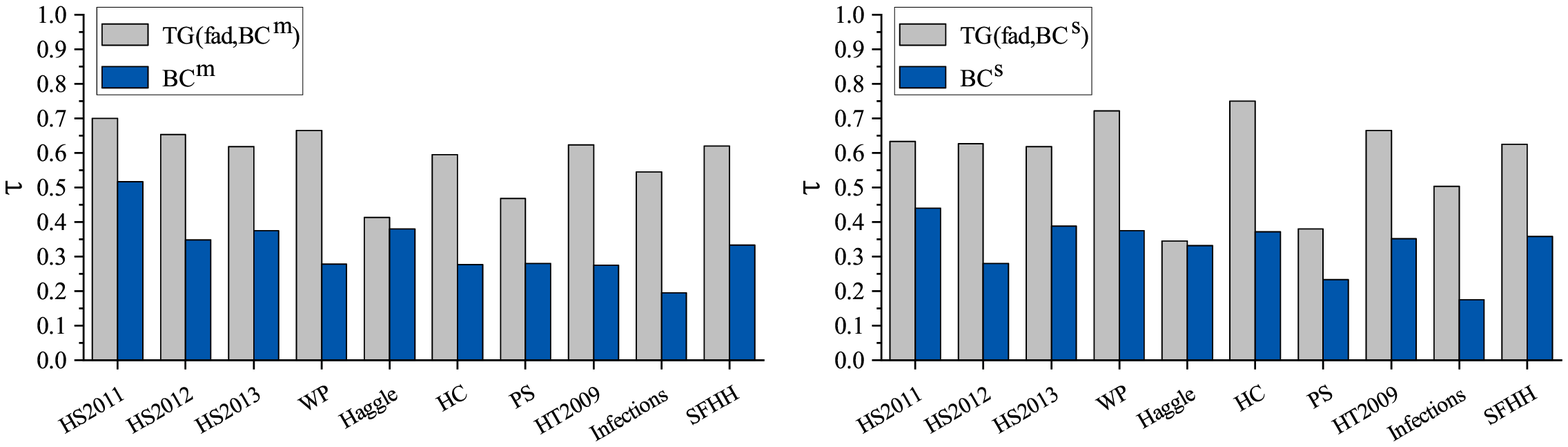}} 
  \caption{The Kendall correlation coefficients $\tau$ between the node importance score derived from $NE_{fad}$ and the centrality metrics. The Kendall correlation coefficients $\tau$ between $NE_{fad}$ and the temporal gravity model are shown in gray histogram, the Kendall correlation coefficients $\tau$ between $NE_{fad}$ and baseline centrality metrics are shown in blue histogram.}
  \label{Figure:4}
\end{figure}

In the temporal gravity model, we use baseline centrality metrics as the node mass. Taking PageRank centrality as an example, the temporal gravity model \textit{$TG(fad,{PR}^m)$} integrates the PageRank score of nearby nodes that are within a certain temporal distance as the centrality value of the target node. The PageRank centrality ${PR}^m$ can also be independently used as a centrality metric. In Figure~\ref{Figure:4}, we show how the temporal gravity model can help to improve the node importance identification compared to that of only using a baseline centrality metric. The baseline centrality metrics are $PR^m, DC^m, CC^m, BC^m$ based on temporal network snapshots and $PR^s, DC^s, CC^s, BC^s$ based on the corresponding aggregated static networks. We show that temporal gravity model based on different baseline centrality metrics outperforms that of only using the corresponding baseline centrality metrics in identifying important nodes across all the network datasets.

\begin{figure}[H]
  \centering
  \subfigure{\includegraphics[width=\textwidth]{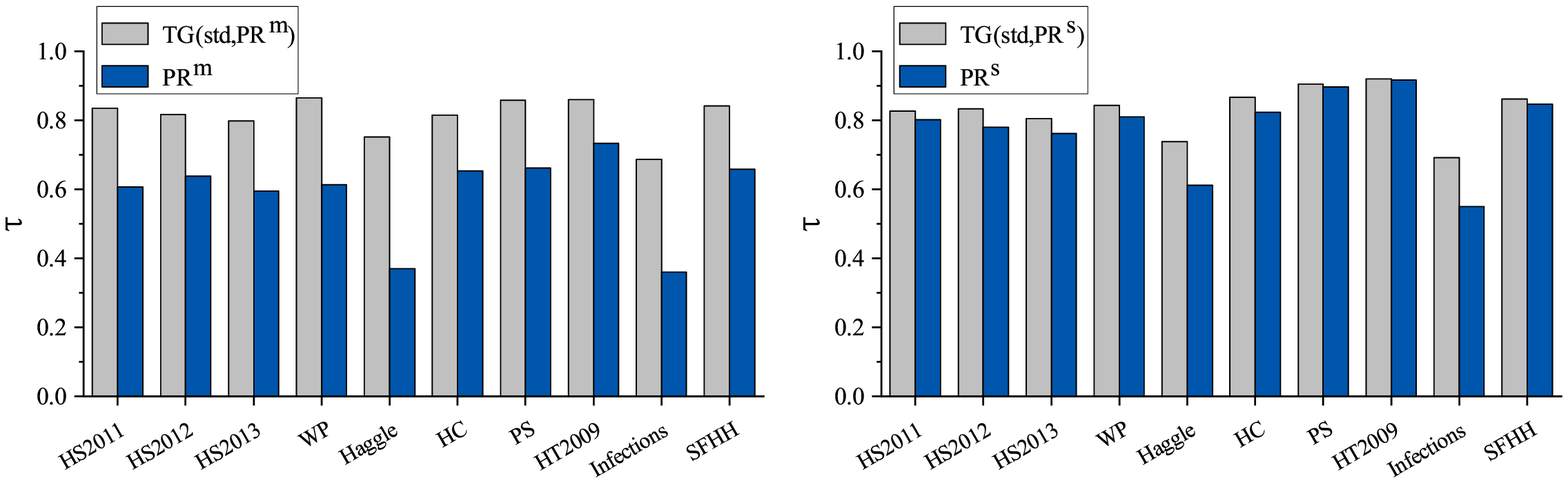}} 
  \subfigure{\includegraphics[width=\textwidth]{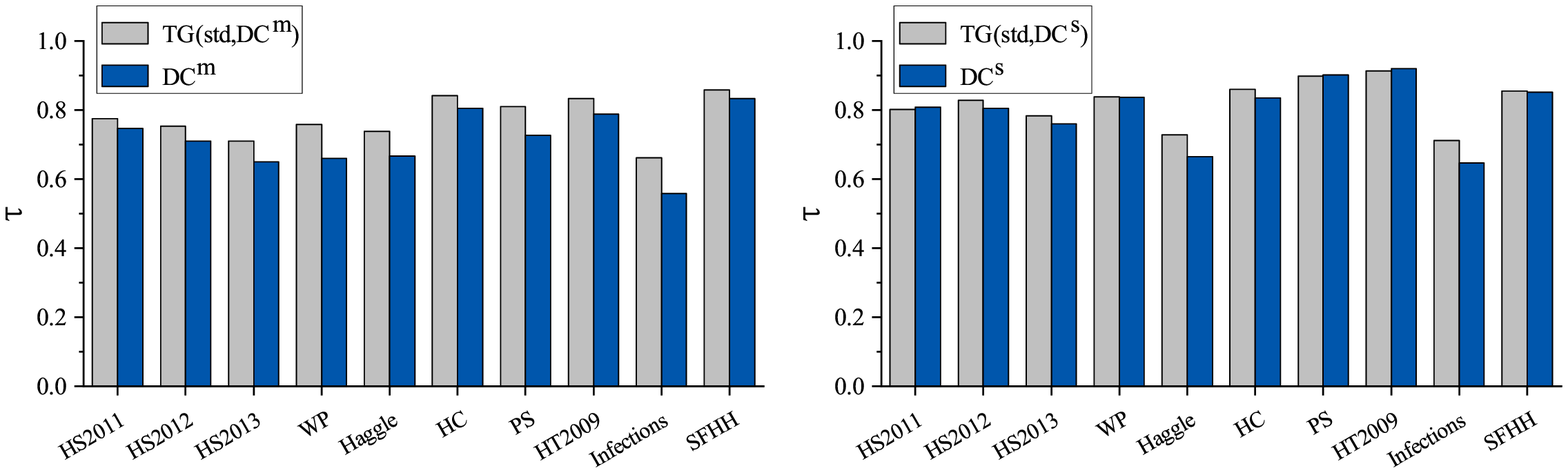}} 
  \subfigure{\includegraphics[width=\textwidth]{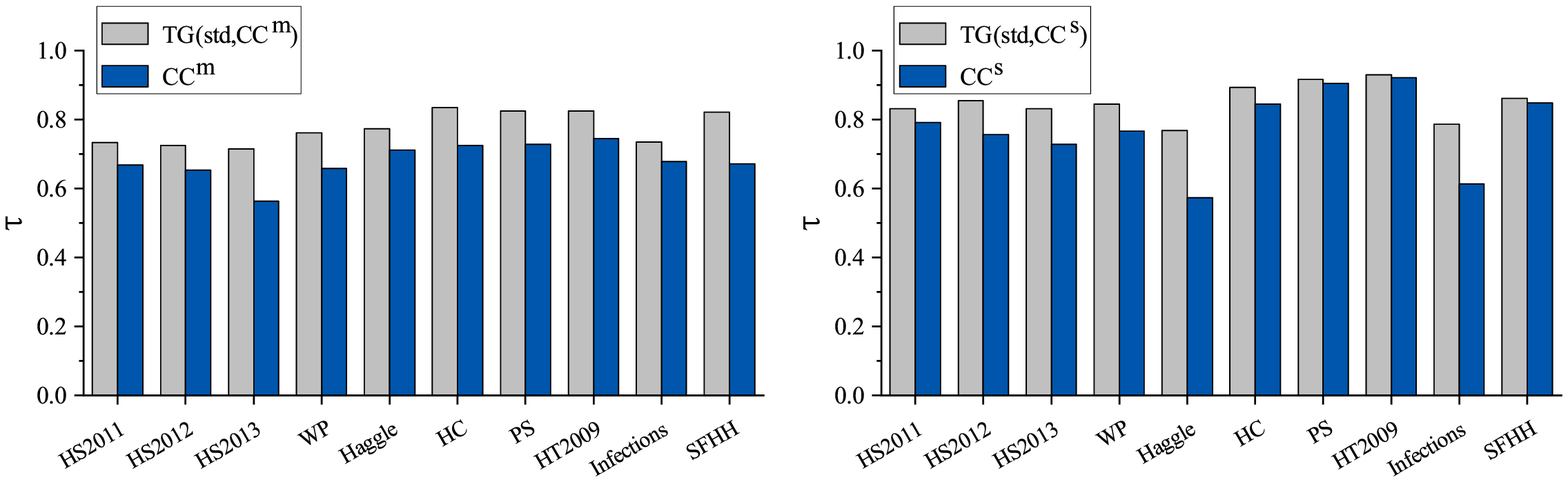}} 
  \subfigure{\includegraphics[width=\textwidth]{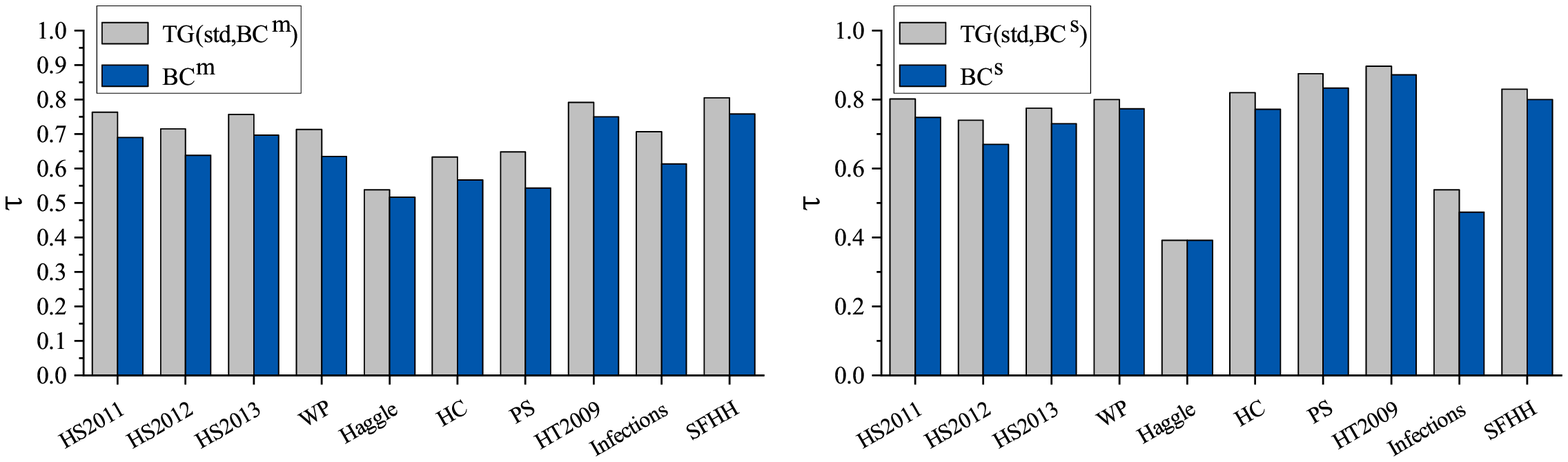}} 
  \caption{The Kendall correlation coefficients $\tau$ between the node importance score derived from $NE_{std}$ and the centrality metrics. The Kendall correlation coefficients $\tau$ between $NE_{std}$ and the temporal gravity model are shown in gray histogram, the Kendall correlation coefficients $\tau$ between $NE_{std}$ and baseline centrality metrics are shown in blue histogram.}
  \label{Figure:3}
\end{figure}

\clearpage
The results of Table~\ref{Table:6} and Figure~\ref{Figure:4} are based on the fastest arrival distance. We further show the results of important node identification based on the temporal shortest distance in Table~\ref{Table:5} and Figure~\ref{Figure:3}. Due to the high computational complexity of the temporal shortest distance, we consider the truncation radius $R=5$. We obtain the similar results as that of using the fastest arrival distance in Table~\ref{Table:6} and Figure~\ref{Figure:4}.

\begin{table}[H]
\caption{The Kendall correlation coefficient $\tau$ between the the node centrality score derived from $NE_{std}$ and the centrality metrics for ten empirical networks. The best $\tau$ of each network data is emphasized in bold and asterisk. The best $\tau$ of each network data derived from the baseline metrics is emphasized in bold.}
\centering
\resizebox{\textwidth}{!}{
\begin{tabular}{|l|r|r|r|r|r|r|r|r|r|r|}
\hline
\rowcolor[HTML]{9B9B9B}
   & \multicolumn{1}{l|}{\cellcolor[HTML]{9B9B9B}\textbf{HS2011}}    & \multicolumn{1}{l|}{\cellcolor[HTML]{9B9B9B}\textbf{HS2012}}    & \multicolumn{1}{l|}{\cellcolor[HTML]{9B9B9B}\textbf{HS2013}}    & \multicolumn{1}{l|}{\cellcolor[HTML]{9B9B9B}\textbf{WP}}        & \multicolumn{1}{l|}{\cellcolor[HTML]{9B9B9B}\textbf{Haggle}}    & \multicolumn{1}{l|}{\cellcolor[HTML]{9B9B9B}\textbf{HC}}        & \multicolumn{1}{l|}{\cellcolor[HTML]{9B9B9B}\textbf{PS}}        & \multicolumn{1}{l|}{\cellcolor[HTML]{9B9B9B}\textbf{HT2009}}    & \multicolumn{1}{l|}{\cellcolor[HTML]{9B9B9B}\textbf{Infections}} & \multicolumn{1}{l|}{\cellcolor[HTML]{9B9B9B}\textbf{SFHH}}      \\ \hline
$TG(std,DD)$  & 0.78438                                                                           & 0.82806                                                                           & 0.80627                                                                           & 0.85509                        & 0.76777                        & \textbf{0.92072$^\ast$}                                                     & 0.90810                                                                       & 0.91972                                                                       & 0.77978                         & 0.84791                        \\ \hline
$TG(std,TD)$  & 0.82908                                                                           & 0.84606                                                                           & 0.82162                                                                           & \textbf{0.86895$^\ast$} & 0.77002                        & 0.90270                                                                            & \textbf{0.92024$^\ast$}                                                & 0.92541                                                                       & 0.76933                         & 0.85707                        \\ \hline
$TG(std,{PR}^m)$  & \textbf{0.83568$^\ast$}                                                    & 0.81664                                                                           & 0.79850                                                                           & 0.86465                        & 0.75202                        & 0.81477                                                                            & 0.85878                                                                       & 0.85936                                                                       & 0.68646                         & 0.84245                        \\ \hline
$TG(std,{PR}^s)$ & 0.82603                                                                           & 0.83389                                                                           & 0.80473                                                                           & 0.84410                        & 0.73863                        & 0.86595                                                                            & 0.90576                                                                       & 0.92004                                                                       & 0.69142                         & \textbf{0.86102$^\ast$} \\ \hline
$TG(std,{DC}^m)$  & 0.77575                                                                           & 0.75394                                                                           & 0.70935                                                                           & 0.75905                        & 0.73772                        & 0.84144                                                                            & 0.80975                                                                       & 0.83281                                                                       & 0.66215                         & 0.85865                        \\ \hline
$TG(std,{DC}^s)$ & 0.80190                                                                           & 0.82893                                                                           & 0.78342                                                                           & 0.83885                        & 0.72857                        & 0.86018                                                                            & 0.89897                                                                       & 0.91308                                                                       & 0.71088                         & 0.85578                        \\ \hline
$TG(std,{CC}^m)$ & 0.73410                                                                           & 0.72551                                                                           & 0.71569                                                                           & 0.76192                        & \textbf{0.77269$^\ast$} & 0.83568                                                                            & 0.82497                                                                       & 0.82491                                                                       & 0.73450                         & 0.82129                        \\ \hline
$TG(std,{CC}^s)$ & 0.83238                                                                           & \textbf{0.85475$^\ast$}                                                   & \textbf{0.83227$^\ast$}                                                    & 0.84458                        & 0.76787                        & 0.89333                                                                            & 0.91729                                                                       & \textbf{0.93047$^\ast$}                                                & \textbf{0.78681$^\ast$} & 0.86102                        \\ \hline

$TG(std,{BC}^m)$   & 0.76348 & 0.71455 & 0.75699 & 0.71398 & 0.53780 & 0.63315 & 0.64768 & 0.79235 & 0.70658 & 0.80424  \\ \hline
$TG(std,{BC}^s)$  & 0.80165 & 0.73950 & 0.77516 & 0.80062 & 0.39244 & 0.82054 & 0.87428 & 0.89697 & 0.53774 & 0.82952  \\ \hline

${PR}^m$  & 0.60686                                                                           & 0.63811                                                                           & 0.59509                                                                           & 0.61331                        & 0.36954                        & 0.65261                                                                            & 0.66249                                                                       & 0.73357                                                                       & 0.35927                         & 0.65796                        \\ \hline
${DC}^m$ & 0.74688                                                                           & 0.71057                                                                           & 0.65033                                                                           & 0.66043                        & 0.66724                        & 0.80535                                                                            & 0.72641                                                                       & 0.78863                                                                       & 0.55862                         & 0.83356                        \\ \hline
${CC}^m$ & 0.66806                                                                           & 0.65338                                                                           & 0.56309                                                                           & 0.65870                        & \textbf{0.71209} & 0.72541                                                                            & 0.72868                                                                       & 0.74431                                                                       & \textbf{0.67835}  & 0.67209                        \\ \hline
${BC}^m$ & 0.68982                                                                           & 0.63795                                                                           & 0.69643                                                                           & 0.63447                        & 0.51634                        & 0.56685                                                                            & 0.54357                                                                       & 0.74968                                                                       & 0.61397                         & 0.75830                        \\ \hline
${PR}^s$ & 0.80140                                                                           & 0.78001                                                                           & \textbf{0.76136}                                                    & 0.80970                        & 0.61131                        & 0.82342                                                                            & 0.89719                                                                       & 0.91688                                                                       & 0.54921                         & 0.84734                        \\ \hline
${DC}^s$ & \textbf{0.80797}                                                   & \textbf{0.80506}                                                    & 0.75958                                                                           & \textbf{0.83620}& 0.66532                        & 0.83427                                                                            & 0.90135                                                                       & 0.92011                                                                       & 0.64727                         & \textbf{0.85200} \\ \hline
${CC}^s$ & 0.79243                                                                           & 0.75621                                                                           & 0.72848                                                                           & 0.76588                        & 0.57406                        & \textbf{0.84466}                                                     & \textbf{0.90571 }                                               & \textbf{0.92223 }                                               & 0.61324                         & 0.84834                        \\ \hline
${BC}^s$ & 0.74908                                                                           & 0.66960                                                                           & 0.72980                                                                           & 0.77338                        & 0.39181                        & 0.77153                                                                            & 0.83265                                                                       & 0.87137                                                                       & 0.47303                         & 0.79981                        \\ \hline
\end{tabular}}
\label{Table:5}
\end{table}

\subsection{Performance Evaluation Based on Spreading Capacity}
\begin{figure}[H]
\centering
\includegraphics[width=0.5\textwidth]{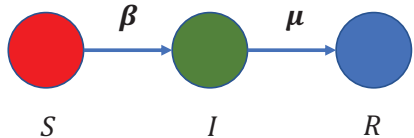}
\caption{Schematic diagram of the SIR spreading model.}
\label{SIR}
\end{figure}

A node is influential if a piece of information starts from it can spread to a large population. We call a seed node that can result in a large spreading size as a node with high spreading capacity. In this section, we evaluate the performance of our temporal gravity model and the baseline metrics in identifying influential nodes on temporal networks. We use the SIR model to simulate the spreading process on temporal networks~\cite{qu2019temporal, zhan2019information}. In the SIR spreading model, a node can be in one of the following three states, i.e., susceptible (S), infected (I), and recovered (R). We show a schematic diagram of SIR model in Figure~\ref{SIR}. A susceptible node can be infected by an infected node with probability $\beta$ if there is a contact between them. An infected node can transfer to the recovered state with probability $\mu$. The spreading process follows the time flow of the temporal networks. In the following experiments, the infection probability is fixed as $\beta=0.1$ and recovery probability is fixed as $\mu=0.01$.

In a temporal network with $n$ snapshots $G_1,G_2,\cdots, G_n$, a node can appear in multiple snapshots. Therefore, if we choose a node as the seed of the spreading process, we need to consider which time to start the spreading process. For a node $v_i$, we assume the time when it appears is given by $T_{v_{i}}=\{ t^1_{v_i}, t^2_{v_i}, \cdots, t^m_{v_i}\}$. We take every time step $t^j_{v_i} \in T_{v_{i}}$ as the starting time of the spreading process. For node $v_i$ as the seed and $t^j_{v_i}$ as the start time, we run the SIR spreading process until the end of the temporal network to get the final spreading range $R^j_{v_i}$. The final spreading range contains the infected and recovered nodes at the final state of the spreading process. For each $t^j_{v_i}$ as the starting time, we run the spreading process for $100$ times to get the final average spreading range as $\overline{R^j_{v_i}}$. For each seed node ${v_i}$, we run the spreading process starting at every occurrence time $t^j_{v_i}$ for $100$ times. Therefore, the spreading ranges are recorded as $R(v_i)=\{\overline{R^1_{v_i}},\overline{R^2_{v_i}}, \cdots, \overline{R^{m}_{v_i}}\}$. We illustrate three ways to define a node's spreading capacity. The max spreading capacity of node ${v}_i$ is the largest spreading range over all its occurrence time,
donated as $R_{max}(v_i)=\max\{\overline{R^j_{v_i}}|\overline{R^j_{v_i}}\in R(v_i)\}$.
The average spreading capacity of node ${v}_i$ is the average spreading range over the set $R(v_i)$, donated as $R_{mean}(v_i)$. The normalized spreading capacity is donated as 
$R_{norm}(v_i)=\frac{1}{m}\sum_{j=1}^{m}\frac{\overline{R^j_{v_i}}}{n-t^j_{v_i}+1}$. 
A node with larger value of $R_{max}$, $R_{mean}$ or $R_{norm}$ implies the node has larger spreading capacity.

We evaluate the temporal gravity model and the baseline centrality metrics in identifying nodes that have high spreading capacity in temporal networks. The temporal gravity model used here is based on the temporal shortest distance. The real temporal capacity of a node is defined by $R_{max}$, $R_{mean}$ or $R_{norm}$. Taking temporal gravity model as an example, we illustrate how to use Kendall correlation coefficient to evaluate its performance. First of all, we compute the node importance score for every node by the temporal gravity model to get a list of centrality score of every node. Then we run the spreading process illustrated above to get the spreading capacity list for every node. The Kendall correlation coefficient $\tau$ is computed between the centrality score list and the spreading capacity list. A high  value of $\tau$ indicates the centrality metric can better identify influential nodes.

\begin{figure}[H]
  \centering
  \subfigure{\includegraphics[width=\textwidth]{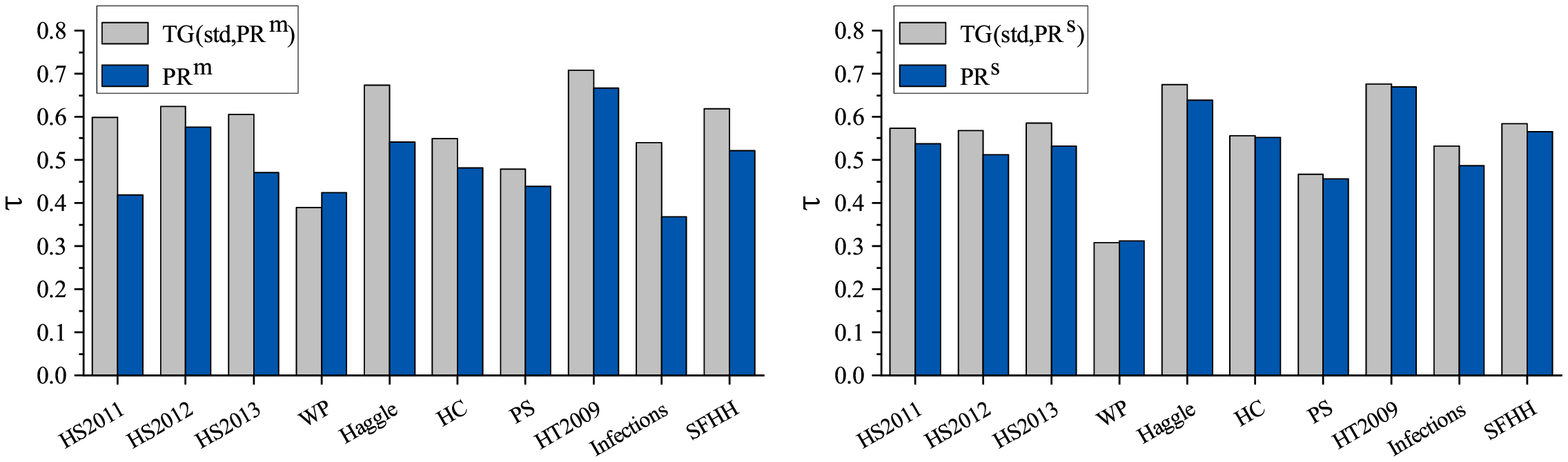}} 

  \subfigure{\includegraphics[width=\textwidth]{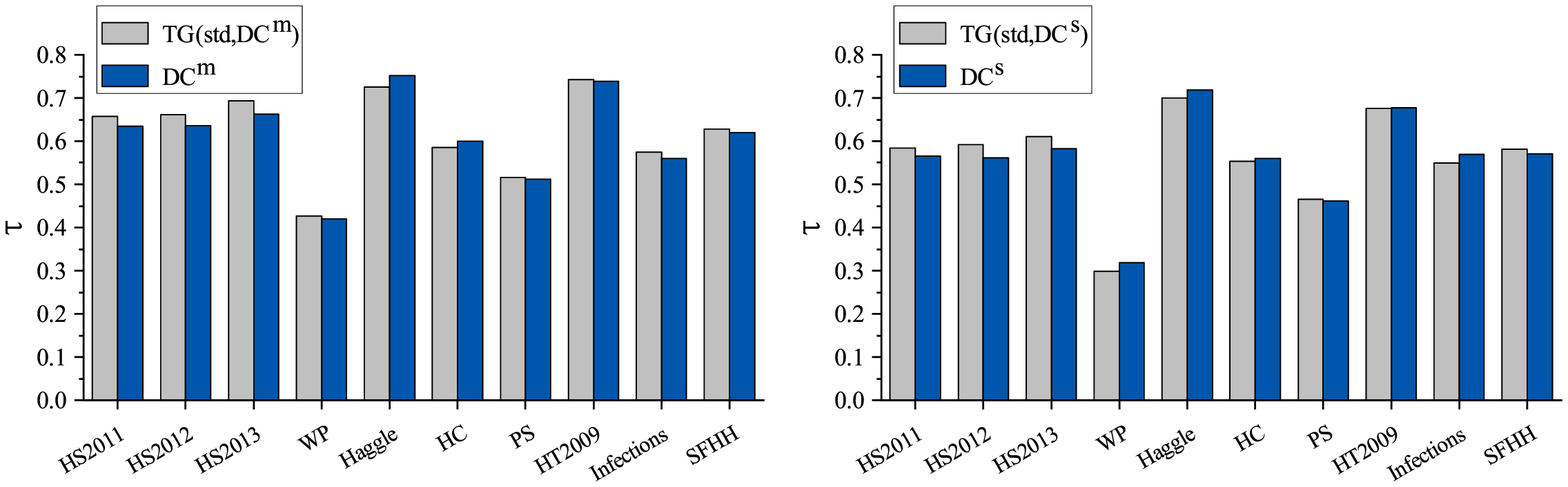}} 

  \subfigure{\includegraphics[width=\textwidth]{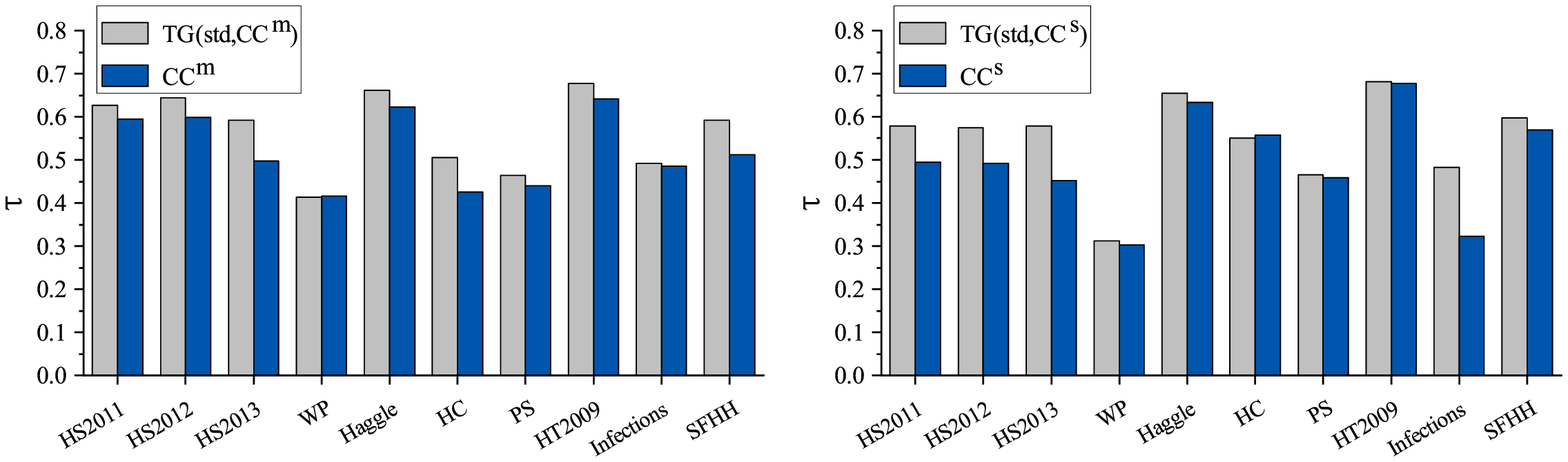}} 

  \subfigure{\includegraphics[width=\textwidth]{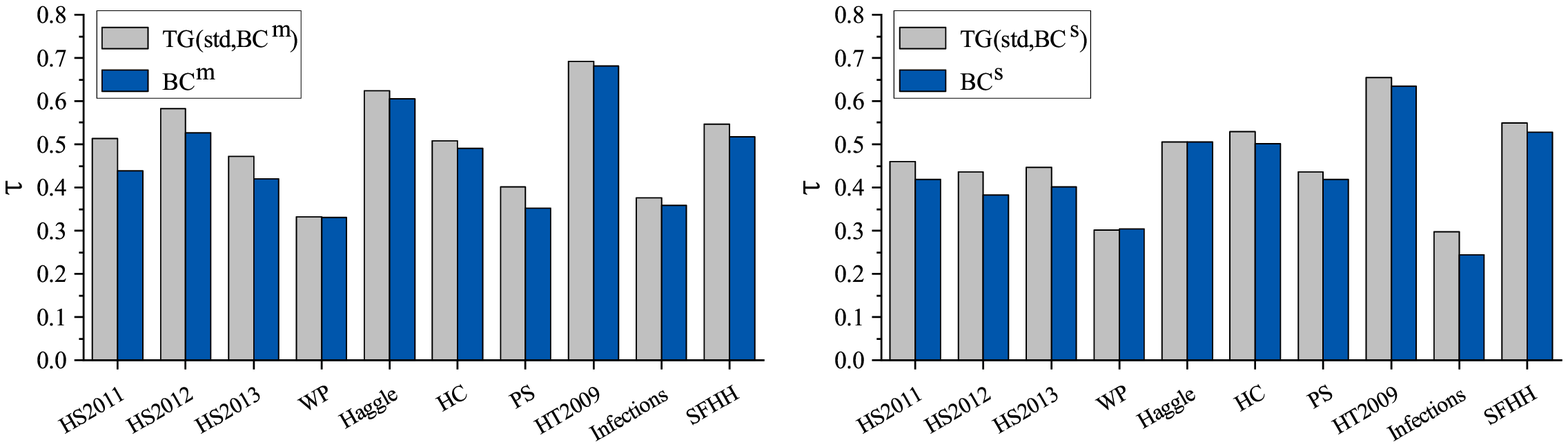}} 
  \caption{The Kendall correlation coefficients $\tau$ between node real spreading capacity  $R_{norm}$  and  node centrality score derived the centrality metrics. The Kendall correlation coefficients $\tau$ between $R_{norm}$ and the temporal gravity model are shown in gray histogram, the Kendall correlation coefficients $\tau$ between $R_{norm}$ and baseline centrality metrics are shown in blue histogram.}
  \label{Figure:5}
\end{figure}

We perform the centrality metrics on identifying influential nodes, the results are show in Table~\ref{Table:7}, Table 8 (see SI), Table 9 (see SI) for spreading capacity defined by $R_{norm}$, $R_{max}$, $R_{mean}$, respectively. The results show that the temporal gravity model performs better in identifying influential nodes in the majority of the temporal network datasets. Even for some datasets that temporal gravity model doesn't outperform, Kendall correlation coefficients can maintain a higher level. In Figure~\ref{Figure:5}, we show the temporal gravity model based on baseline centrality metrics performs better than only using baseline centrality metrics in identifying influential nodes.

\begin{table}[H]
\caption{The Kendall correlation coefficient $\tau$ between node real spreading capacity  $R_{norm}$  and  node centrality score derived the centrality metrics for ten empirical networks. The best $\tau$ of each network data is emphasized in bold and asterisk. The best $\tau$ of each network data derived from the baseline metrics is emphasized in bold.}
\centering
\resizebox{\textwidth}{!}{
\begin{tabular}{|l|r|r|r|r|r|r|r|r|r|r|}
\hline
\rowcolor[HTML]{9B9B9B}
   & \multicolumn{1}{l|}{\cellcolor[HTML]{9B9B9B}\textbf{HS2011}}    & \multicolumn{1}{l|}{\cellcolor[HTML]{9B9B9B}\textbf{HS2012}}    & \multicolumn{1}{l|}{\cellcolor[HTML]{9B9B9B}\textbf{HS2013}}    & \multicolumn{1}{l|}{\cellcolor[HTML]{9B9B9B}\textbf{WP}}        & \multicolumn{1}{l|}{\cellcolor[HTML]{9B9B9B}\textbf{Haggle}}    & \multicolumn{1}{l|}{\cellcolor[HTML]{9B9B9B}\textbf{HC}}        & \multicolumn{1}{l|}{\cellcolor[HTML]{9B9B9B}\textbf{PS}}        & \multicolumn{1}{l|}{\cellcolor[HTML]{9B9B9B}\textbf{HT2009}}    & \multicolumn{1}{l|}{\cellcolor[HTML]{9B9B9B}\textbf{Infections}} & \multicolumn{1}{l|}{\cellcolor[HTML]{9B9B9B}\textbf{SFHH}}      \\ \hline
$TG(std,DD)$                    & 0.59975 & 0.60348 & 0.61648 & 0.31008 & 0.65338 & 0.56757 & 0.46572 & 0.68805 & 0.48945 & 0.60061 \\ \hline
$TG(std,TD)$                    & 0.59467 & 0.57529 & 0.60053 & 0.32394 & 0.64551 & 0.56252 & 0.46607 & 0.68363 & 0.51049 & 0.60142 \\ \hline
$TG(std,{PR}^m)$ & 0.59924                        & 0.62359                        & 0.60481                        & 0.38939                        & 0.67388                        & 0.54955                        & 0.47882                        & 0.70796                        & 0.54029                        & 0.61821                        \\ \hline
$TG(std,{PR}^s)$  & 0.57283                        & 0.56760                        & 0.58575                        & 0.30769                        & 0.67522                        & 0.55604                        & 0.46655                        & 0.67636                        & 0.53137                        & 0.58379                        \\ \hline
$TG(std,{DC}^m)$  & \textbf{0.65714$^\ast$} & \textbf{0.66096$^\ast$} & \textbf{0.69359$^\ast$} & \textbf{0.42714$^\ast$} & 0.72548                        & 0.58486                        & \textbf{0.51620$^\ast$} & \textbf{0.74210$^\ast$} & \textbf{0.57452$^\ast$} & \textbf{0.62799$^\ast$} \\ \hline
$TG(std,{DC}^s)$  & 0.58425                        & 0.59243                        & 0.61126                        & 0.29861                        & 0.69979                        & 0.55315                        & 0.46552                        & 0.67573                        & 0.54868                        & 0.58192                        \\ \hline
$TG(std,{CC}^m)$ & 0.62717                        & 0.64444                        & 0.59149                        & 0.41376                        & 0.66152                        & 0.50559                        & 0.46408                        & 0.67794                        & 0.49205                        & 0.59157                        \\ \hline
$TG(std,{CC}^s)$ & 0.57917                        & 0.57480                        & 0.57854                        & 0.31199                        & 0.65413                        & 0.55027                        & 0.46518                        & 0.68173                        & 0.48330                        & 0.59718                        \\ \hline
$TG(std,{BC}^m)$& 0.51381 & 0.58320 & 0.47167 & 0.33186 & 0.62426 & 0.50775 & 0.40146 & 0.69216 & 0.37630 & 0.54656 \\ \hline
$TG(std,{BC}^s)$ & 0.46057 & 0.43633 & 0.44639 & 0.30148 & 0.50590 & 0.52937 & 0.43658 & 0.65518 & 0.29761 & 0.54942 \\ \hline
${PR}^m$ & 0.41867                        & 0.57666                        & 0.47059                        & \textbf{0.42379 }& 0.54135                        & 0.48108                        & 0.43918                        & 0.66688                        & 0.36852                        & 0.52167                        \\ \hline
${DC}^m$ & \textbf{0.63467} & \textbf{0.63588} & \textbf{0.66289} & 0.41990                        & \textbf{0.75201$^\ast$} & \textbf{0.60004$^\ast$} & \textbf{0.51150} & \textbf{0.73921} & 0.55985                        & \textbf{0.61998 }\\ \hline
${CC}^m$  & 0.59416                        & 0.59814                        & 0.49674                        & 0.41567                        & 0.62246                        & 0.42559                        & 0.44062                        & 0.64096                        & 0.48547                        & 0.51259                        \\ \hline
${BC}^m$ & 0.43812                        & 0.52671                        & 0.42033                        & 0.33058                        & 0.60559                        & 0.49045                        & 0.35181                        & 0.68173                        & 0.35857                        & 0.51717                        \\ \hline
${PR}^s$ & 0.53752                        & 0.51173                        & 0.53172                        & 0.31151                        & 0.63826                        & 0.55243                        & 0.45551                        & 0.66941                        & 0.48628                        & 0.56537                        \\ \hline
${DC}^s$ & 0.56493                        & 0.56181                        & 0.58323                        & 0.31907                        & 0.71907                        & 0.55945                        & 0.46196                        & 0.67751                        & \textbf{0.56921} & 0.57046                        \\ \hline
${CC}^s$ & 0.49522                        & 0.49158                        & 0.45245                        & 0.30237                        & 0.63317                        & 0.55753                        & 0.45885                        & 0.67698                        & 0.32204                        & 0.56927                        \\ \hline
${BC}^s$ & 0.41816                        & 0.38331                        & 0.40110                        & 0.30387                        & 0.50545                        & 0.50198                        & 0.41868                        & 0.63401                        & 0.24377                        & 0.52846                        \\ \hline
\end{tabular}}
\label{Table:7}
\end{table}

\section{Discussions and Conclusions}
In real world, most complex systems are dynamic. To preserve the temporal information of the systems, we generally represent them as temporal networks. Although many centrality metrics have been proposed for static networks, important node identification on temporal networks is still an open challenge.

The gravity law is a simple, elegant and representative formula, which estimates the strength of the interaction between two objects by considering both the inherent influence of the two objects and the distance between them. Inspired by the idea of the gravity law and the success of previous proposed gravity models on static networks, we propose a temporal gravity model to identify important nodes in temporal networks. The temporal gravity model takes advantage of both the neighborhood information and the temporal information in temporal networks. Also, it provides a mathematical and computational framework which can includes different node properties as the node masses and different temporal distance as the distance in the gravity formula. For node property, we use node's baseline centrality metrics as well as two types of temporal degree, i.e., time degree (TD), distance degree (DD).  For the temporal distance, we consider two cases, i.e., the fastest arrival distance and the temporal shortest distance.

The Kendall correlation is an index measuring the ranking correlation between two variables. We use the Kendall correlation coefficient $\tau$ between a node’s real influence on the network connectivity (or spreading process) and the importance score derived from different centrality metrics to evaluate the performance of the centrality metrics on temporal empirical networks. For network efficiency, we consider two types: network efficiency based on the temporal shortest path $NE_{std}$ and based on the fastest arrival path $NE_{fad}$. For the $NE_{fad}$, we find that $TG$-$fad$ methods performs better than the baseline centrality metrics across different networks. The Kendall correlation coefficient $\tau$ of temporal gravity model increases by 105.31$\%$ on average compared to the best value of $\tau$ derived from baseline centrality metrics across the ten empirical networks. For the $NE_{std}$, the $TG$-$std$ methods have an overall improvement. The $TG$-$std$ methods based on $DD$ and $TD$ show steady improvement over baseline centrality metrics. Regarding to the spreading influence, we simulate SIR spreading model on the empirical temporal networks. We use the normalized, maximal and the average spreading range to represent a node's spreading capacity, respectively. The $TG$-$std$ methods based on the baseline centrality metrics ("Averaged") outperforms the baseline centrality metrics. On the whole, temporal gravity model shows robust node identification performance across networks. Temporal gravity model using baseline centrality metrics as it mass has an overall improvement over the corresponding baseline centrality metrics. Our model has superior robustness for important node identification on temporal networks. Since the success of the gravity model in important node identification on both static and temporal networks, future work can try to extend it to other networks, such as multi-layer networks~\cite{kivela2014multilayer,boccaletti2014structure} and bipartite networks~\cite{newman2018networks}. Also, the spreading influence of a node can be different if we use different dynamical processes. One can further explore how to identify influential nodes for spreading processes, such as susceptible-infected-susceptible (SIS)~\cite{zhan2018coupling} and coevolution spreading process~\cite{wang2019coevolution}, on a temporal network.

\clearpage
\section{Appendix}
We evaluate the temporal gravity model and the baseline centrality metrics in identifying nodes that have high spreading capacity in temporal networks. The following are the results based on the maximal spreading range and average spreading range.

\indent Maximal spreading range ${R}_{max}$:
\begin{table}[H]
\caption{The Kendall's tau correlation coefficients between $R_{max}$ and temporal network methods for ten empirical networks. The best performed method for each network is emphasized in bold and asterisk. The best performed traditional method for each network is emphasized in bold. For the SIR model, parameters are $\beta$=0.1, $\mu$=0.01. For each node at one occurrence time, the spreading range result is the average of 100 times independently simulations.}
\centering
\resizebox{\textwidth}{!}{
\begin{tabular}{|l|r|r|r|r|r|r|r|r|r|r|}
\hline
\rowcolor[HTML]{9B9B9B}
   & \multicolumn{1}{l|}{\cellcolor[HTML]{9B9B9B}\textbf{HS2011}}    & \multicolumn{1}{l|}{\cellcolor[HTML]{9B9B9B}\textbf{HS2012}}    & \multicolumn{1}{l|}{\cellcolor[HTML]{9B9B9B}\textbf{HS2013}}    & \multicolumn{1}{l|}{\cellcolor[HTML]{9B9B9B}\textbf{WP}}        & \multicolumn{1}{l|}{\cellcolor[HTML]{9B9B9B}\textbf{Haggle}}    & \multicolumn{1}{l|}{\cellcolor[HTML]{9B9B9B}\textbf{HC}}        & \multicolumn{1}{l|}{\cellcolor[HTML]{9B9B9B}\textbf{PS}}        & \multicolumn{1}{l|}{\cellcolor[HTML]{9B9B9B}\textbf{HT2009}}    & \multicolumn{1}{l|}{\cellcolor[HTML]{9B9B9B}\textbf{Infections}} & \multicolumn{1}{l|}{\cellcolor[HTML]{9B9B9B}\textbf{SFHH}}      \\ \hline
$TG(std,DD)$                    & 0.71293                        & 0.58452                        & 0.58919                        & 0.59259                        & 0.66063                        & 0.60743                        & 0.37767                        & 0.57016                        & 0.72069                         & 0.48165                        \\ \hline
$TG(std,DD)$                      & 0.66366                        & 0.56763                        & 0.57384                        & 0.59594                        & 0.65201                        & 0.60815                        & 0.38427                        & 0.56321                        & 0.70514                         & 0.48380                        \\ \hline
$TG(std,{PR}^m)$   & 0.61794                        & 0.60761                        & 0.57774                        & 0.62413                        & 0.68070                        & 0.58940                        & 0.42730                        & 0.57174                        & 0.62891                         & 0.51471                        \\ \hline
$TG(std,{PR}^s)$   & 0.62353                        & 0.54082                        & 0.55080                        & 0.58925                        & 0.68354                        & 0.61031                        & 0.37842                        & 0.54393                        & 0.65180                         & 0.46350                        \\ \hline
$TG(std,{DC}^m)$ & 0.70074                        & 0.63977                        & \textbf{0.63860$^\ast$} & 0.64851                        & 0.72910                        & \textbf{0.61464$^\ast$} & 0.42936                        & 0.60714                        & 0.71411                         & 0.51205                        \\ \hline
$TG(std,{DC}^s)$ & 0.65071                        & 0.55671                        & 0.56348                        & 0.58495                        & 0.71378                        & 0.61031                        & 0.37540                        & 0.54267                        & 0.69244                         & 0.45940                        \\ \hline
$TG(std,{CC}^m)$ & \textbf{0.73376$^\ast$} &  \textbf{0.64138$^\ast$} & 0.60059                        & \textbf{0.67240$^\ast$} & 0.67358                        & 0.57859                        & 0.38983                        & 0.61315                        & \textbf{0.72220$^\ast$}  & \textbf{0.52422$^\ast$} \\ \hline
$TG(std,{CC}^s)$ & 0.64969                        & 0.54603                        & 0.55095                        & 0.58495                        & 0.66095                        & 0.60743                        & 0.38021                        & 0.55563                        & 0.70075                         & 0.47906                        \\ \hline
$TG(std,{BC}^m)$ & 0.55639                        & 0.55579                        & 0.44170                        & 0.58860                        & 0.64000                        & 0.50577                        & 0.33505                        & 0.54330                        & 0.59608                         & 0.43884                        \\ \hline
$TG(std,{BC}^s)$ & 0.47927                        & 0.42866                        & 0.43248                        & 0.53047                        & 0.53874                        & 0.60238                        & 0.38516                        & 0.53540                        & 0.38150                         & 0.44804                        \\ \hline
${PR}^m$          & 0.42517                        & 0.56391                        & 0.46059                        & 0.58017                        & 0.54639                        & 0.50793                        & \textbf{0.45548$^\ast$} & 0.53635                        & 0.32578                         & 0.47832                        \\ \hline
${DC}^m$           & 0.62412                        & \textbf{0.62970} & \textbf{0.61048} & 0.62038                        & \textbf{0.76118$^\ast$} & \textbf{0.60966} & 0.45392                        & 0.60806                        & 0.64310                         & \textbf{0.51162} \\ \hline
${CC}^m$          & \textbf{0.67280} & 0.62710                        & 0.53661                        & \textbf{0.63990} & 0.64140                        & 0.51947                        & 0.39368                        & \textbf{0.62231$^\ast$} & \textbf{0.69643}  & 0.51106                        \\ \hline
${BC}^m$          & 0.42517                        & 0.56391                        & 0.46059                        & 0.58017                        & 0.54639                        & 0.50793                        & 0.45548                        & 0.53635                        & 0.32578                         & 0.47832                        \\ \hline
${PR}^s$           & 0.53438                        & 0.49066                        & 0.49789                        & 0.58638                        & 0.64842                        & 0.60526                        & 0.38124                        & 0.53192                        & 0.51626                         & 0.45044                        \\ \hline
${DC}^s$          & 0.58925                        & 0.52341                        & 0.53304                        & 0.60355                        & 0.74947                        & 0.60911                        & 0.38060                        & 0.53615                        & 0.64440                         & 0.45150                        \\ \hline
${CC}^s$         & 0.53828                        & 0.43828                        & 0.43354                        & 0.52349                        & 0.63889                        & 0.60282                        & 0.38145                        & 0.54003                        & 0.51027                         & 0.45322                        \\ \hline
${BC}^s$            & 0.43228                        & 0.36471                        & 0.39117                        & 0.51756                        & 0.53810                        & 0.60382                        & 0.38234                        & 0.51233                        & 0.30332                         & 0.43458                        \\ \hline
\end{tabular}}
\label{Table:8}
\end{table}

\indent Average spreading range ${R}_{mean}$:

\begin{table}[H]
\caption{The Kendall's tau correlation coefficients between $R_{mean}$ and temporal network methods for ten empirical networks. The best performed method for each network is emphasized in bold and asterisk. The best performed traditional method for each network is emphasized in bold. For the SIR model, parameters are $\beta$=0.1, $\mu$=0.01. For each node at one occurrence time, the spreading range result is the average of 100 times independently simulations.}
\centering
\resizebox{\textwidth}{!}{
\begin{tabular}{|l|r|r|r|r|r|r|r|r|r|r|}
\hline
\rowcolor[HTML]{9B9B9B}
   & \multicolumn{1}{l|}{\cellcolor[HTML]{9B9B9B}\textbf{HS2011}}    & \multicolumn{1}{l|}{\cellcolor[HTML]{9B9B9B}\textbf{HS2012}}    & \multicolumn{1}{l|}{\cellcolor[HTML]{9B9B9B}\textbf{HS2013}}    & \multicolumn{1}{l|}{\cellcolor[HTML]{9B9B9B}\textbf{WP}}        & \multicolumn{1}{l|}{\cellcolor[HTML]{9B9B9B}\textbf{Haggle}}    & \multicolumn{1}{l|}{\cellcolor[HTML]{9B9B9B}\textbf{HC}}        & \multicolumn{1}{l|}{\cellcolor[HTML]{9B9B9B}\textbf{PS}}        & \multicolumn{1}{l|}{\cellcolor[HTML]{9B9B9B}\textbf{HT2009}}    & \multicolumn{1}{l|}{\cellcolor[HTML]{9B9B9B}\textbf{Infections}} & \multicolumn{1}{l|}{\cellcolor[HTML]{9B9B9B}\textbf{SFHH}}      \\ \hline
$TG(std,DD)$                    & 0.68686                        & 0.52948                        & 0.49393                        & 0.49307                        & 0.58228                        & 0.38739                        & 0.25572                        & 0.50411                        & \textbf{0.69346$^\ast$}  & 0.44590                        \\ \hline
$TG(std,TD)$                      & 0.63911                        & 0.51421                        & 0.47716                        & 0.50024                        & 0.57419                        & 0.38667                        & 0.26072                        & 0.49905                        & 0.67884                         & 0.44923                        \\ \hline
$TG(std,{PR}^m)$   & 0.58324                        & 0.54364                        & 0.47205                        & 0.55996                        & 0.59817                        & 0.38234                        & 0.28830                        & 0.50569                        & 0.60043                         & 0.46883                        \\ \hline
$TG(std,{PR}^s)$   & 0.60356                        & 0.48939                        & 0.46147                        & 0.48782                        & 0.60385                        & 0.39604                        & 0.26710                        & 0.48546                        & 0.63600                         & 0.43293                        \\ \hline
$TG(std,{DC}^m)$ & 0.65486                        & \textbf{0.58150$^\ast$} & \textbf{0.54140$^\ast$} & 0.63497                        & 0.64153                        & \textbf{0.40613$^\ast$} & 0.28452                        & 0.52465                        & 0.67760                         & 0.47096                        \\ \hline
$TG(std,{DC}^s)$ & 0.63276                        & 0.50726                        & 0.47881                        & 0.48065                        & 0.63543                        & 0.39892                        & 0.26100                        & 0.48609                        & 0.66925                         & 0.43066                        \\ \hline
$TG(std,{CC}^m)$ & \textbf{0.68737$^\ast$} & 0.57418                        & 0.48740                        & 0.63306                        & 0.59234                        & 0.37441                        & 0.25188                        & \textbf{0.53445$^\ast$} & 0.68208                         & \textbf{0.47380$^\ast$} \\ \hline
$TG(std,{CC}^s)$ & 0.63429                        & 0.49460                        & 0.45772                        & 0.47874                        & 0.58313                        & 0.39171                        & 0.26024                        & 0.49210                        & 0.67898                         & 0.44513                        \\ \hline
$TG(std,{BC}^m)$ & 0.52118                        & 0.52410                        & 0.35925                        & 0.51105                        & 0.58798                        & 0.32036                        & 0.20805                        & 0.47977                        & 0.59671                         & 0.41199                        \\ \hline
$TG(std,{BC}^s)$ & 0.46489                        & 0.36655                        & 0.35056                        & 0.41949                        & 0.49752                        & 0.38523                        & 0.29131                        & 0.47440                        & 0.39009                         & 0.41302                        \\ \hline
${PR}^m$        & 0.37625                        & 0.49472                        & 0.34519                        & 0.59245                        & 0.47045                        & 0.34847                        & \textbf{0.29296$^\ast$} & 0.45828                        & 0.29588                         & 0.41575                        \\ \hline
${DC}^m$           & 0.57652                        & \textbf{0.57501} & \textbf{0.51372} & 0.63273                        & 0.65827                        & \textbf{0.40557} & 0.29069                        & 0.51620                        & 0.60895                         & \textbf{0.46900} \\ \hline
${CC}^m$          & \textbf{0.61879} & 0.55742                        & 0.42230                        & \textbf{0.64262$^\ast$} & 0.55577                        & 0.33766                        & 0.24145                        & \textbf{0.52339} & \textbf{0.64752}  & 0.44884                        \\ \hline
${BC}^m$           & 0.43279                        & 0.48946                        & 0.31272                        & 0.48835                        & 0.57167                        & 0.31027                        & 0.18597                        & 0.46492                        & 0.51072                         & 0.39357                        \\ \hline
${PR}^s$           & 0.51390                        & 0.44022                        & 0.41712                        & 0.50215                        & 0.57457                        & 0.39243                        & 0.27348                        & 0.47724                        & 0.50993                         & 0.42064                        \\ \hline
${DC}^s$           & 0.57084                        & 0.47402                        & 0.45397                        & 0.50700                        & \textbf{0.66973$^\ast$} & 0.39369                        & 0.26684                        & 0.48043                        & 0.62944                         & 0.42280                        \\ \hline
${CC}^s$          & 0.54144                        & 0.39203                        & 0.35006                        & 0.40685                        & 0.57901                        & 0.38962                        & 0.26424                        & 0.48079                        & 0.51059                         & 0.42174                        \\ \hline
${BC}^s$          & 0.41994                        & 0.30460                        & 0.31135                        & 0.41328                        & 0.49707                        & 0.38955                        & 0.28795                        & 0.45575                        & 0.31524                         & 0.40204                        \\ \hline
\end{tabular}}
\label{Table:9}
\end{table}

\section*{Declaration of interests}
\textbf{Jialin Bi}: Conceptualization, Methodology, Software, Formal analysis,Writing - Original Draft, Writing - Review \& Editing, Visualization. \textbf{Ji Jin}: Software, Validation, Formal analysis, Investigation, Data Curation. \textbf{Cunquan Qu}: Methodology, Writing - Original Draft, Writing - Review \& Editing, Supervision. \textbf{Xiuxiu Zhan}: Conceptualization, Methodology, Investigation, Writing - Review \& Editing. \textbf{Guanghui Wang}: Resources, Writing - Review \& Editing, Supervision, Funding acquisition.

\section*{Declaration of interests}
The authors declare that they have no known competing financial interests or personal relationships that could have appeared to influence the work reported in this paper.

\section*{Acknowledgements}
 We thank the SocioPatterns collaboration (http://www.sociopatterns.org) and the KONECT (http://konect.uni-koblenz.de/networks) for providing the data sets. The scientific calculations in this paper have been done on the HPC Cloud Platform of Shandong University

\bibliography{myref}

\end{document}